\documentclass[12pt]{article}

\usepackage{amsmath,amsfonts,braket,algorithm,graphicx,multirow,amsthm}
\usepackage{fullpage,authblk}

\theoremstyle{definition} 
\theoremstyle{definition} 
\newtheorem {theorem} {Theorem}

\newcommand{\kb}[1]{\ket{#1}\bra{#1}}

\newcommand{\bk}[1]{\braket{#1|#1}}

\newcommand{\MR}{\texttt{Measure and Resend}}
\newcommand{\R}{\texttt{Reflect}}
\newcommand{\M}{\texttt{Measure}}

\newcommand{\Perm}{\texttt{Permute}}

\newcommand{\prf}{p^{A\rightarrow B}}
\newcommand{\prr}{p^{A\rightarrow A}}

\newcommand{\tdl}{\left|\left|}
\newcommand{\tdr}{\right|\right|}

\floatname{algorithm}{Protocol}

\newcounter{OpenProblemCount}
\newcommand{\openproblem}[0]{\textbf{Open Problem \arabic{OpenProblemCount}: }\stepcounter{OpenProblemCount}}

\newcommand{\protocol}[1]{$ $\newline\newline\textbf{Protocol: #1}}
\newcommand{\subheading}[1]{$ $\newline\textbf{#1:}}

\title{Semi-Quantum Cryptography}
\author[1]{Hasan Iqbal}
\author[1]{Walter O. Krawec\footnote{Email: \texttt{walter.krawec@uconn.edu}}}
\affil[1]{\small{Department of Computer Science and Engineering}\\\small{University of Connecticut}\\\small{Storrs, CT 06269 USA}}

\begin{document}
\maketitle
\begin{abstract}
Quantum key distribution (QKD) protocols allow two parties to establish a shared secret key, secure against an all powerful adversary.  This is a task impossible to achieve through classical communication only; indeed, to distribute a secret key through classical means requires one to assume  computationally bounded adversaries.  If, however, both parties are ``quantum capable'' then security may be attained assuming only that the adversary must obey the laws of physics.  But ``how quantum'' must a protocol actually be to gain this advantage over classical communication?  This is one of the questions semi-quantum cryptography seeks to answer.

Semi-quantum communication, a model introduced in 2007 by M. Boyer, D. Kenigsberg, and T. Mor (PRL 99 140501), involves the use of fully-quantum users and semi-quantum, or ``classical'' users.  These classical users are only allowed to interact with the quantum channel in a limited, classical manner.  Originally introduced to study the key-distribution problem, semi-quantum research has since expanded, and continues to grow, with new protocols, security proof methods, experimental implementations, and new cryptographic applications beyond key distribution.  Research in the field of semi-quantum cryptography requires new insights into working with restricted protocols and, so, the tools and techniques derived in this field can translate to results in broader quantum information science.  Furthermore, other questions such as the connection between quantum and classical processing, including how classical information processing can be used to counteract a quantum deficiency in a protocol, can shed light on important theoretical questions.

This work surveys the history and current state-of-the-art in semi-quantum research.  We discuss the model and several protocols offering the reader insight into how protocols are constructed in this realm.  We discuss security proof methods and how classical post-processing can be used to counteract users' inability to perform certain quantum operations.  Moving beyond key distribution, we survey current work in other semi-quantum cryptographic protocols and current trends.  We also survey recent work done in attempting to construct practical semi-quantum systems including recent experimental results in this field.  Finally, as this is still a growing field, we highlight, throughout this survey, several open problems that we feel are important to investigate in the hopes that this will spur even more research in this topic.
\end{abstract}

\section{Introduction}
Through most of history, \emph{cryptography} was an \emph{art} primarily focused on hiding and sending information secretly (i.e., encryption).  Numerous ciphers were used through history, many of which are now considered insecure by modern standards.  In fact, it wasn't until very recently in the mid 20th century that cryptography transformed from an art to a \emph{science}.  Now we have rigorous methods and techniques to argue and measure security of cryptographic systems.  Interestingly, along with these new techniques came a great extension to the underlying applications beyond encryption, such as authentication, secret sharing, signatures (just to name a few), along with a great explosion in user base.

In general, there are two flavors to modern cryptography: \emph{private key} and \emph{public key} (also known as \emph{symmetric key} and \emph{asymmetric key} respectively).  In a private key setting, all users of the underlying primitive (whether it be encryption, authentication, or some other task), share the same secret key $k$ (i.e., this key is privately shared and all users have the same symmetric information concerning this key).  In a public key setting, one user has a \emph{public/private} key pair while all other users, including potential adversaries, hold the public key.  Thus there is an asymmetry to the overall system with one user having additional information.

While public key cryptography is an incredibly useful mechanism, allowing for users with only public information to, for instance, send information securely to a single party holding the secret key (namely, public key encryption), it is also orders of magnitude slower than symmetric key systems.  Furthermore, while some symmetric key systems can be proven information theoretic secure (i.e., secure without requiring computational assumptions), this is impossible with public key cryptography where security must \emph{always} depend on some unproven computational assumption.  Thus, despite public key cryptography's great appeal, it is still desirable in practice to use symmetric key cryptography whenever possible.  But how do parties agree on a secret key $k$ without an adversary learning it?  This is exactly the \emph{key distribution} problem.  For more information on these issues, the reader is referred to \cite{katz-crypto}.

Of course, if distances are short and the user-base is small, a secret key could be agreed on by meeting in person.  Obviously this solution does not scale.  Today, we use public key cryptography to distribute secret keys (for instance in a TLS/SSL connection, public key encryption is used to transfer a randomly generated shared session key between parties \cite{stallings2007network} - thus, public key cryptography is used to ``boot-strap'' symmetric key mechanisms which are much faster).  But the security of such systems, then, depend entirely on the security of the underlying public key system used to distribute the key.

Rather interestingly, it is a mathematical impossibility for two parties to agree on a shared secret key which is secure against a computationally unbounded adversary, when using only \emph{classical communication}.  Instead, one must always make assumptions on the power of the adversary.  Thankfully, this impossibility result does not hold when parties switch to \emph{quantum communication}.  Indeed, if users utilize quantum information (in addition to classical information), it is possible to establish a shared secret key, secure against an all-powerful adversary (i.e., an adversary who is bounded only by the laws of physics, and not necessarily by some computational hardness assumption).  Requiring only that the adversary live in a physical universe, as opposed to the adversary having difficulty solving certain mathematical problems, is an arguably safer assumption for securing our communication infrastructure.

\emph{Quantum Key Distribution} (QKD) was initially discovered in 1984 by Bennett and Brassard \cite{QKD-BB84} and, independently, in 1991 by Ekert \cite{QKD-E91}, however it wasn't until many years later in 2001 that a full proof of security was developed \cite{QKD-BB84-rate1}.  An alternative, information theoretic proof technique was developed in 2004 by Renner et al. \cite{QKD-renner-keyrate}.  Such protocols require the two users, whom we refer throughout this work as the customary Alice ($A$) and Bob ($B$), to both be capable of manipulating qubits in certain manners (e.g., being able to prepare and measure qubits in arbitrary bases).  Both parties must, therefore, be \emph{quantum capable}.

But is this always needed?  If both parties are capable only of classical communication, perfectly secure key distribution is impossible; if both parties are quantum capable, then it is possible.  What is the middle-ground and what exactly happens in this ``gap'' between classical and quantum communication?  This is the question which \emph{semi-quantum cryptography} seeks to shed light on.  Introduced originally in 2007 by M. Boyer, D. Kenigsberg, and T. Mor in \cite{SQKD-first}, this field has seen growing interest over the years with new protocols, new cryptographic primitives, and new security proofs leading to a growing research area.  Furthermore, as our society begins to move towards practical implementations of quantum communication networks \cite{qkd-survey,qkd-survey2,qkd-survey3,qkd-survey4}, the semi-quantum model may hold unique benefits allowing for potentially cheaper devices (as less ``quantum capable hardware'' may be required) or devices that are more robust to hardware faults (as one may switch to a semi-quantum mode of operation if some devices fail).  Finally, the theoretical and practical innovations necessary to study the semi-quantum model, where users are highly restricted in their abilities, may lead to great innovations in the broader field of quantum information science.

This paper surveys the development and the latest state of the art in the field of semi-quantum cryptography.  We will begin by discussing basic (fully) quantum key distribution topics that are relevant.  After this, we will present the semi-quantum model in detail along with the first protocols developed for key distribution - so called \emph{semi-quantum key distribution} (SQKD) protocols.  We will cover in detail semi-quantum key distribution protocols, past and current along with the research being done to reduce quantum resource requirements even further.  Following this, a detailed review of security results and methods will be presented, including proof techniques and noise tolerance results.  The last topic in key-distribution will be a survey of multi-user SQKD protocols, including protocols where multiple classical users establish a key through the use of an untrusted (adversarial) quantum server.

Semi-quantum cryptography now spreads beyond key distribution so we will then survey other applications of the model to primitives such as secret sharing, secure direct communication, and private state comparison.  We will conclude with a survey on current practical SQKD research, including recent experimental implementations.

\subsection{Quantum Key Distribution}

Before discussing the semi-quantum model of cryptography, we review here some basic facts about quantum key distribution.  We only review some important facts needed to put into context the work done in the semi-quantum model - for a more complete survey of standard (i.e., ``fully-quantum'') quantum cryptographic protocols and technology, the reader is referred to \cite{qkd-survey,qkd-survey2,qkd-survey3,qkd-survey4}.  This survey, of course, will focus on semi-quantum communication and cryptography.

QKD protocols utilize both quantum and classical communication.  A quantum channel connects users allowing them to send quantum resources to one another (e.g., qubits); an authenticated classical channel is also available on which users may send authenticated, but not secret, messages to one another.  A QKD protocol generally consists of two stages: first is a \emph{quantum communication stage} followed by a second \emph{post processing stage}.  Much research is often spent in the first stage, while the second generally consists of standard classical cryptographic processes (though, we note, developing new, faster, and more efficient systems for this second stage is also an area of active research and interest).  Certainly, in the semi-quantum field, at the moment all research has been on the first stage, using standard techniques and methods for the second stage.

The quantum communication stage of a protocol typically operates over numerous, independent, iterations.  Each iteration consists of random choices by users, however the choice is independent of previous iterations.  Thus, when presenting a protocol later, we generally write out only a single iteration of the quantum communication stage - \emph{it is understood, then, that what is written would be repeated a sufficiently large amount of time as required by the users}.  The goal of this stage is to utilize the quantum communication channel and the classical authenticated channel, to establish what is called a \emph{raw key}.  These are two strings of classical bits, one string held by $A$ and one by $B$, which are partially correlated and partially secret.  Due to noise in the quantum channel (either natural or adversarially generated), the strings will inevitably have errors in them.  Furthermore, one must assume the worst case that an adversary has some classical or quantum system correlated or entangled with this raw key.  Thus, the raw key, by itself, cannot be used directly for cryptographic applications.  Instead it must be further processed through the second stage of a QKD protocol.  Before this, however, a second output of the quantum communication stage is some form of sampling data on the quantum channel determining, at a minimum, the noise level in the quantum channel.  We refer to this data as the channel's \emph{noise signature}.  Exactly what data this consists of depends on the protocol.

The second stage, the classical post processing stage, takes as input the raw key and the noise signature and, first, runs an error correction protocol.  This uses the authenticated classical channel thus leaking additional information to $E$; this additional leakage must be taken into account in any security proof.  Following this a \emph{privacy amplification} protocol is run, taking the error corrected raw key and hashing it down to a secret key.  Privacy amplification is done using a two-universal hash function \cite{QKD-renner-keyrate}.  Namely, $A$ will choose a hash function $f$ randomly from a family of two-universal hash functions.  She will send a description of $f$ to $B$ using the authenticated channel (thus the adversary then knows which function $f$ is used).  Following this, both users run their raw key through the hash function resulting in a secret key $K_A$ and $K_B$ of size $\ell$ bits.

If the protocol is correct, it should hold that $K_A = K_B$ with high probability (i.e., they should differ only with negligible probability as determined by some user specified security parameter).  If the protocol is secure, it should be that any adversary's system should be independent of the final secret key and, furthermore, the secret key should be no different from one chosen uniformly at random.

More formally, let $\rho_{KE}$ be the state of the system describing the generated secret key $K$ (known to $A$ and $B$) and $E$'s system (which includes anything learned from error correction and the chosen privacy amplification hash function).  Then, the protocol is considered secure if:
\begin{equation}\label{eq:qkd-secure}
\tdl\rho_{KE} - I_K/2^\ell\otimes\rho_E\tdr \le \epsilon,
\end{equation}
where $\tdl X\tdr$ is the trace distance of operator $X$.  In essence, the above says that, after execution, the actual protocol state, $\rho_{KE}$, is $\epsilon$-indistinguishable from an ideal state consisting of a key chosen uniformly at random and completely independent of $E$'s system.  One is very often interested in the \emph{key rate} of a (S)QKD protocol, defined to be the ratio of secret bits ($\ell$) to either the size of the raw-key $N$, or the number of qubit signals sent (the latter of which is, of course, never smaller than $N$ and can, in fact, be much larger depending on how efficient the protocol is); note that the latter term is smaller and is often called the \emph{effective key-rate}.  We will return to these notions later when we discuss key-rate computations for SQKD protocols.

As we will see later, an important question, given a new (S)QKD protocol, is to determine a bound on its key-rate $\ell/N$, either in the finite key setting or the asymptotic setting (the latter being when $N$ approaches infinity).  This bound should be a function only of observed noise statistics (the noise signature).  Once this is computed, one is also interested in a (S)QKD protocol's \emph{noise tolerance} - namely the maximal observed noise for which the key-rate remains positive.  We will return to these concepts later.

\section{Semi-Quantum Key Distribution}

The question, ``how quantum does a protocol or system need to be to gain an advantage over its classical counterpart'' is an important one both theoretically and practically.  This question has been studied in various manners, but it wasn't until Boyer et al., introduced the semi-quantum model for key distribution that this question was first extended to the field of \emph{cryptography} \cite{SQKD-first}.

A semi-quantum key distribution (SQKD) protocol typically consists of two users: a \emph{fully quantum user} Alice ($A$) and a \emph{classical user} Bob ($B$) (though the names may occasionally be reversed in some references, it is irrelevant to our discussion).  Before introducing actual protocols, it is important to more rigorously understand and define exactly what a ``classical'' user is in the context of this problem.  Indeed, what does it even mean for a so-called ``classical'' user to interact with a quantum channel?

\subsection{The Semi-Quantum Model}

\begin{figure}
  \centering
  \includegraphics[width=250pt]{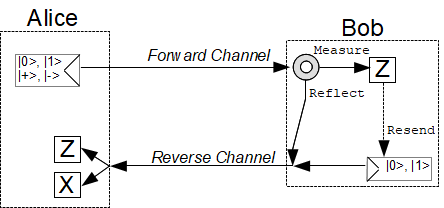}
\caption{The typical setup of an SQKD protocol.  A fully quantum user $A$ begins by sending quantum states to $B$.  This user, $A$, is allowed to prepare any state required by the protocol.  This qubit passes through the \emph{forward channel} to the semi-quantum or classical user $B$.  This user, $B$, is only allowed to perform certain operations, namely ignore the qubit ($\R$) or interact with the qubit in the computational $Z$ basis only (e.g., $\MR$).  $B$ may also only prepare $Z$ basis qubits.  The \emph{reverse channel} then carries qubits back to $A$.  Note that, $B$, the classical user, can only interact with the channel directly in the $Z$ basis, or disconnect from the channel, in which case $A$ is sending quantum information to herself in a large loop.  When qubits return to $A$ through the reverse channel, she is allowed to perform any operation on them.}\label{fig:sqkd-diagram}
\end{figure}

The quantum user has access to a quantum channel which starts at her lab, travels out, and returns to her.  The \emph{classical user} (sometimes called the \emph{semi-quantum user}) $B$ can access a portion of this channel.  Thus, semi-quantum protocols operate over a two-way quantum channel where qubits, or other quantum carriers, travel first from the quantum user $A$, to the classical user $B$, then return to the quantum user.  See Figure \ref{fig:sqkd-diagram}.

Besides the requirement of a two-way quantum channel, SQKD protocols place a further restriction on the classical user.  Namely, he is only allowed to interact with the quantum channel by performing a $Z$ basis measurement or sending $Z$ basis qubits.  Alternatively, he can simply ignore the channel, disconnect from it, in which case $A$, the quantum user, is simply ``talking to herself.''  More specifically, for every qubit received, $B$ may choose one of the following options:
\begin{enumerate}
  \item $\M$: Subject the incoming qubit to a $Z$ basis measurement.
  \item $\texttt{Prepare}$: Prepare a $Z$ basis state and send it to $A$ on the reverse channel.
  \item $\MR$: Subject the incoming qubit to a $Z$ basis measurement and then resend the result back to $A$ as a $Z$ basis qubit (a combination of the above two operations, though with the restriction that $B$ always sends the same state he observes - this is actually important for security as discussed later in this paper).
  \item $\R$: Reflect the qubit back to $A$ undisturbed.  This is equivalent to simply disconnecting from the quantum channel and forwarding everything back to the sender (e.g., $A$).  The classical user does not learn anything about the state of the qubit in this case.
  \item $\Perm$: Reorder the qubits received (or a subset of qubits received) without otherwise disturbing them.  The classical user does not learn anything about the underlying states of the qubits, he only permutes their order.
\end{enumerate}

Protocols that utilize $\R$ and $\MR$ are generally called \emph{measure-resend} protocols while those that require $\Perm$ are usually called \emph{randomization based} protocols \cite{sqkd-second}.  Note that $B$ can only directly work in the $Z$ basis; otherwise he can disconnect from the quantum channel in which case $A$, the fully quantum user, is simply ``talking to herself.''  We comment that, occasionally, new semi-quantum research introduces new operations that the semi-quantum user $B$ may perform and we will comment on these other operations as they arise in our survey.

Note that if both $A$ and $B$ were restricted to these operations (either working in a single $Z$ basis or disconnecting from the quantum channel), the resulting protocol would be no different, from a mathematical standpoint, from a purely classical protocol and, thus, could never be unconditionally secure.  The question is, can one user be classical in this sense (in that he can only operate in a single basis or ignore the channel) while still maintaining security?  As it turns out, the answer is, indeed, yes.


\subsection{First Protocols}

In their seminal paper, Boyer et al. \cite{SQKD-first}, introduced two SQKD protocols to demonstrate the model.  Here we discuss these protocols to give the reader some idea as to how SQKD protocols typically operate.

As with QKD protocols, SQKD protocols operate in two stages.  The quantum communication stage of the original measure-resend style SQKD protocol from \cite{SQKD-first}, which we denote here BKM07, operates by repeating the following:
\protocol{BKM07 \cite{SQKD-first}}
\begin{enumerate}
  \item $A$ prepares one of the four states $\ket{0}, \ket{1}, \ket{+}$, or $\ket{-}$ with uniform probability, remembering her choice.  She sends the resulting qubit to the classical user $B$.
  \item $B$ will choose to either $\MR$ or to $\R$ the incoming qubit.
  \item $A$ will measure the returning qubit in the same basis she initially used to prepare it ($Z$ or $X$).
  \item $A$ discloses her choice of basis while $B$ discloses his choice of operation.  Measurement results and initial state preparation choices remain secret.
  \item If $A$ choose to send a $Z$ basis qubit and if $B$ choose $\MR$, parties may use this round for their raw-key.  In particular, $A$ will use her initial preparation choice while $B$ will use his measurement result (these should be correlated).  All other iterations, along with a suitably sized random sample of raw-key iterations, are used for sampling the quantum channel error rates.
\end{enumerate}

Note that the protocol could be written equivalently with $A$ using her measurement result for her key-bit instead of her initial state preparation choice.  Indeed, $A$ can choose later which option to follow based on the observed noise - if the noise in the forward channel is higher than the reverse, it would make sense to switch (this way there will be a higher correlation between $A$ and $B$'s raw key results).  For sampling, note that due to the two way quantum channel, several statistics can be gathered, namely:
\begin{itemize}
  \item $\prf_{i,j}$: The probability that $B$ measures a $\ket{j}$ given that $A$ sent a $\ket{i}$ and $B$ choose $\MR$, for $i, j \in \{0,1\}$.
  \item $p^{B\rightarrow A}_{i,j}$: The probability that $A$ measures a $\ket{j}$ given that $B$ sent a $\ket{i}$ (for $i,j \in \{0,1\}$).
  \item $\prr_{i,R,j}$: The probability that $A$ measures a $\ket{j}$ given that she initially sent a $\ket{i}$ and $B$ choose $\R$ (now, for $i,j \in \{0,1,+,-\}$).
\end{itemize}
Note that, while users can measure the $Z$ basis noise (e.g., a $\ket{i}$ flipping to a $\ket{1-i}$ for $i=0,1$) in each of the forward and reverse channels, they can only measure the $X$ basis noise in the entire joint channel.  Since $B$ cannot measure or prepare in the $X$ basis, it is impossible to observe the $X$ basis error in either channel separately.  This opens up potential attack strategies for an adversary and makes security analyses difficult (we comment on current security techniques later in this paper).

The second SQKD protocol introduced by the same authors in \cite{sqkd-second}, utilized the $\Perm$ operation as opposed to the $\MR$.

\protocol{BGKM09 \cite{sqkd-second}}
\begin{enumerate}
  \item $A$ prepares $N$ qubits, each qubit prepared randomly in one of the four states $\ket{0}, \ket{1}, \ket{+}$, or $\ket{-}$.  She sends all $N$ qubits to $B$.
  \item For each qubit, $B$ chooses randomly to $\M$ or to $\R$.  For those qubits he chooses to $\R$, he also permutes the qubits before returning them.  That is, he does not disturb their state through any measurement, however he re-orders them before resending.  Those qubits he measures he does not resend.
  \item $A$ stores the returning qubits in a quantum memory.  At this point, $B$ will inform her which qubits he choose $\R$ and also the order he reflected them back.  The quantum user $A$ then undoes the permutation and measures the returned qubits in the same basis she initially sent them.
  \item $A$ discloses which qubits she sent in the $Z$ basis.  Whenever $A$ sent a $Z$ basis qubit and $B$ choose to $\M$, users now have a correlation used for their raw key.
\end{enumerate}

Note that the above protocol requires a quantum memory on the part of the quantum user.  It also, in a way, requires more advanced capabilities on the part of the classical user in that he must be able to randomly permute qubits (perhaps, through delay lines \cite{sqkd-second}).  Note that, if $B$ does not ``resend'' then he must ``permute'' otherwise the protocol becomes susceptible to the so-called double CNOT attack \cite{sqkd-second}.  Indeed, in this case, $E$ can apply a CNOT gate to all qubits traveling in the forward direction.  If $B$ measures, but does not resend, $E$ will notice a vacuum leaving his lab in which case a measurement of her ancilla provides full information to her.  If $B$ chooses to $\R$, then $E$ will simply apply a CNOT gate in the reverse channel, undoing the initial state (since $B$'s operation is the identity operation in this event and so the two CNOT gates invert each other) and thus avoid detection.  Therefore, it is vital for any point-to-point SQKD protocol to have the classical user ``resend'' or ``permute.''  

The above protocols were \emph{prepare-and-measure} protocols whereby qubits are prepared and subsequently measured (similar to BB84).  Additionally, \emph{entanglement based} SQKD protocols were also subsequently proposed where the fully quantum user prepares entangled states, sending one particle to the other (classical) user and holding the other locally (similar to an E91 style protocol \cite{QKD-E91}).  For instance, \cite{ent1} proposed two protocols in this line where the quantum user prepares a Bell state, sending a particle to the classical user.  This user then performs the $\R$ or $\MR$ operation, returning the state to $A$.  Whenever $B$ performed $\R$, $A$ will measure the qubit pair in the Bell basis - she should receive the same Bell state she originally prepared.  On other iterations, she measures her qubit in the $Z$ basis, creating a correlation between the two parties that is used as their raw key ($A$'s qubit measurement in the $Z$ basis should match $B$'s).  A similar protocol was described in that same reference where $B$ also uses the $\Perm$ option.

An alternative entanglement based protocol was presented in \cite{ent2}.  Here a different encoding scheme was used in that a key bit of $0$ is encoded by sending a Bell state $\ket{\Phi^+} = \frac{1}{\sqrt{2}}(\ket{00}+\ket{11})$ while a key bit of $1$ is encoded by sending a Bell state $\ket{\Psi^+} = \frac{1}{\sqrt{2}}(\ket{01} + \ket{10})$.  Namely, on each iteration of the protocol $A$ prepares, randomly, a Bell state $\ket{\Phi^+}$ or $\ket{\Psi^+}$; her choice determines her random raw key bit for this iteration.  It is, therefore, the goal of $B$ to guess which Bell state $A$ prepared.  This is done by $A$ sending one particle to $B$, keeping the remaining one private.  $B$ then chooses $\MR$ or $\R$.  Finally, when a qubit returns to $A$, she will measure both particles either in the Bell basis or the computational basis.  Parties then disclose their operations.  When $B$ choose $\R$ and $A$ chose to measure in the Bell basis, she should receive the same outcome that she initially prepared (other outcomes being counted as errors).  For all iterations where $A$ measured in the computational basis, however, she will disclose the measurement result of the qubit she kept private (i.e., she discloses $0$ or $1$ based on the computational basis measurement of the qubit she initially kept private).  This disclosure, combined with $B$'s measurement outcome (\emph{which he keeps private}) allows him to determine which of the two Bell states $A$ prepared, thus allowing him to guess $A$'s raw key bit for that iteration.  Note that for $E$ to guess this also, she would have to know the measurement result of the qubit particle that traveled between parties - however such a measurement on her part would have caused a disturbance in the cases where $B$ choose $\R$.  We discuss general security issues later in this work, however this encoding scheme is interesting in that it allows a classical party to, in a way, determine the result of a Bell state preparation with help from $A$.

\subsection{Reducing Resource Requirements}

Numerous SQKD protocols have been developed, beyond those mentioned in the previous section, each with various advantages, disadvantages, and theoretical interests.  One of the primary theoretical goals of semi-quantum cryptography is to better understand ``how quantum'' a protocol must be to gain an advantage over its classical counterpart; thus, one important research direction in semi-quantum communication is in further reducing the quantum resource requirements on the part of the two users.  This includes both the fully-quantum user, and the semi-quantum user.  As this is a vital area of research in semi-quantum cryptography, we spend some time here surveying the recent progress in this area.

The first result in this line of investigation came in 2009 with a paper by Zou et al., \cite{SQKD-less-than-4}.  In this work it was shown, for the first time, that the fully-quantum user can also have reduced resource requirements.  Namely, five new protocols were proposed.  These protocols required fully-quantum Alice to send only three, two, or even a single state to $B$.  On return, of course, the fully-quantum user must be able to measure in two bases.  In light of their work, one may classify SQKD protocols as $n$-state SQKD protocols where $n$ is the number of states that the quantum user $A$ may choose to prepare.  If $n = 1$ we call the protocol a \emph{single state} protocol; otherwise it is a \emph{multi state} protocol.  Zou et al., \cite{SQKD-less-than-4} presented the first single state protocol along with $2$ and $3$ state protocols.  Note that BKM07 is a $4$-state protocol.

The so-called \emph{single state} SQKD protocol from \cite{SQKD-less-than-4} is of particular interest as it sparked several additional protocols along this line; furthermore, such protocols actually admit certain nice reductions in their security proofs which we will comment on later.  The quantum communication stage of this protocol repeats the following process:

\protocol{Single State SQKD \cite{SQKD-less-than-4}}
\begin{enumerate}
  \item $A$ prepares a single qubit in the state $\ket{+}$ and sends it to $B$.
  \item $B$ chooses randomly to $\MR$ or to $\R$ recording his choice and, if applicable, his measurement outcome.
  \item $A$ chooses to measure in the $Z$ or the $X$ basis randomly.
  \item Users $A$ and $B$ disclose their choices.  If $B$ chose to $\MR$ and if $A$ chose to measure in the $Z$ basis, they should share a correlated bit to be used for their raw key.  If $B$ chose $\R$ and if $A$ chose to measure in the $X$ basis, she should observe outcome $\ket{+}$ and any other outcome is considered an error.
\end{enumerate}

This single state protocol is remarkably simple and demonstrated that the fully quantum user need not have advanced source preparation capabilities.  Rather remarkably, as we comment later, the security properties of this protocol were also shown to be optimistically comparable to certain fully-quantum protocols, at least in the perfect qubit scenario (we will discuss this later when we talk about security of SQKD protocols).

Since Zou et al.'s 2009 paper \cite{SQKD-less-than-4}, several other single-state protocols have been proposed.  In 2014 it was shown that a key need not be distilled from measurement choices, but instead may be distilled from $B$'s \emph{action} \cite{krawec2014restricted}.  While the BKM07 protocol may be considered a semi-quantum version of the BB84 protocol (since all four BB84 states are transmitted on the return channel) and the Single State SQKD protocol may be considered a semi-quantum three-state BB84 \cite{QKD-BB84-three-state,QKD-BB84-three-state-v2} (since only three states, $\ket{0}, \ket{1}$, and $\ket{+}$ are transmitted on the return channel), this new protocol is, in a way, a version of the Extended B92 \cite{QKD-B92-extended} protocol.  This is due to the fact that three states are transmitted on the return channel ($\ket{+}$, $\ket{0}$, and $\ket{1}$) and, furthermore, the encoding scheme is based on alternative basis choice (as determined by $B$'s actual operation) and not based on the qubit state directly.  This single-state protocol operates as follows:

\protocol{Reflection-Based SQKD \cite{krawec2014restricted}}
\begin{enumerate}
  \item $A$ prepares a single qubit in the state $\ket{+}$, sending it to $B$.
  \item $B$ chooses a random bit $k_B$.  If $k_B=0$, he will choose to $\R$ the qubit, and furthermore he sets a private internal register $\texttt{accept}_B$ to $\texttt{TRUE}$ with $1/2$ probability (otherwise it is set to $\texttt{FALSE}$).  If $k_B =1$ he chooses to $\MR$ setting $\texttt{accept}_B = \texttt{TRUE}$ only if he observes $\ket{0}$.
  \item $A$ chooses randomly to measure in the $Z$ or $X$ basis.  If she chooses the $Z$ basis \emph{and} observes outcome $\ket{1}$, she sets $k_A = 0$ and $\texttt{accept}_A = \texttt{TRUE}$.  If she chose the $X$ basis and observes outcome $\ket{-}$, she sets $k_A = 1$ and $\texttt{accept}_A = \texttt{TRUE}$.  All other measurement outcome possibilities result in her setting $\texttt{accept}_A = \texttt{FALSE}$ (in which case $k_A$ is set arbitrarily).
  \item Users $A$ and $B$ both divulge their value of $\texttt{accept}_A$ and $\texttt{accept}_B$ respectively.  If both are $\texttt{TRUE}$, they will keep their bits $k_A$ and $k_B$ to contribute towards their raw key.  Otherwise, the iteration is discarded.
\end{enumerate}

It is not difficult to see the similarity between the above protocol and a B92-style protocol.  Indeed, the key is transmitted only when $B$ chooses $\R$ (which should result in him sending a $\ket{+}$) or when he chooses $\MR$ and observes $\ket{0}$.  However, there is one very significant difference - namely, $B$ cannot be certain he is sending a $\ket{+}$ when he chooses $\R$. Indeed, an adversary will attack the forward channel (see Figure \ref{fig:sqkd-diagram}), altering this state.  Unlike security proofs of standard one-way protocols, security proofs of semi-quantum protocols must take into account that what $B$ sends is affected by $E$'s forward channel attack.  This complicates security analyses.  Another protocol based also on this reflection-based encoding scheme was developed in \cite{sqkd-single-state-b92} which is, in a way, the semi-quantum version of B92 (the non-extended version).

In \cite{cl-A}, a \emph{two state} SQKD protocol was proposed where the quantum user prepares a random $X$ basis state (either $\ket{+}$ or $\ket{-}$) and sends it to the classical user.  Note that in their paper, they referred to the quantum user as $B$ and the classical user as $A$ (thus flipping the labels with $B$ now initiating the communication).  However, to remain consistent throughout this work, we will maintain the notion that $A$ is the quantum user who initiates the communication and $B$ is the classical user.  Obviously the exact labeling is irrelevant.  Following this state preparation, the classical user chooses one of two operations $\MR$ or $\R$; finally the quantum user will choose to measure in a random basis $Z$ or $X$.  Both parties disclose their choices and the key is distilled from those iterations where $B$, the classical user, chose $\MR$ and the quantum user $A$ chose the $Z$ basis.  Thus, it is, in a way, a two-state version of the BKM07 protocol.



Another single-state protocol, along with a new four-state protocol, was proposed in \cite{refresh}.  This paper increased $B$'s abilities by allowing him to choose a \texttt{Measure} and \texttt{Prepare} option (as opposed to simply resending whatever he observed).  This option gives $B$ the ability to measure a qubit in the $Z$ basis, but prepare any $Z$ basis state he likes, regardless of his measurement outcome (normally the $\MR$ option forces him to always send the basis state he measured).  This augmentation allows for the construction, also, of a secure direct communication protocol.

\subheading{Attack on $B$'s Send Operation}
Rather interestingly, it was recently shown in \cite{double-cnot} that by allowing $B$ this extra ability (namely the ability to prepare any $Z$ basis state regardless of measurement outcome), only partial security may be achieved.  Thus, rather interestingly, if both parties are fully quantum, secure protocols exists (e.g., BB84 \cite{QKD-BB84}); if $B$ is classical in that he only chooses $\R$, $\MR$, or $\Perm$, the protocol may also be secure (e.g., BKM07 \cite{SQKD-first}); however if we take this and add a little extra power to $B$, security may break down.  The attack discovered in \cite{double-cnot} operates as discussed below, though we generalize it slightly here to show how it may be applied to arbitrary protocols which operate using the $\M$ and $\texttt{Prepare}$ operation (where the prepared $Z$ basis qubit may be in a different state than what was measured) as opposed to the $\MR$ operation - that is, their attack described in \cite{double-cnot} can be applied to arbitrary semi-quantum protocols whenever $B$ deviates from sending exactly the $Z$ basis state he measured:

\begin{enumerate}
  \item In the forward channel (refer to Figure \ref{fig:sqkd-diagram}), Eve applies a CNOT gate, using the traveling qubit as a control and her private ancilla as a target.  Her private ancilla is initially in a $\ket{0}_E$ state.
  \item In the reverse channel, $E$ applies the following operator which acts on the traveling qubit (denoted the $T$ space) and her private ancilla as follows:
\begin{align*}
U_R\ket{00}_{TE} &= \ket{00}_{TE}\\
U_R\ket{11}_{TE} &= \ket{10}_{TE}\\
U_R\ket{01}_{TE} &= \ket{02}_{TE}\\
U_R\ket{10}_{TE} &= \ket{13}_{TE}.
\end{align*}
Note that we assume $E$'s ancilla is four dimensional, spanned by basis states $\ket{0}$, $\cdots$, $\ket{3}$.  Also note that the operator $U_R$ as described is an isometry and so may be dilated to a unitary operator through standard techniques (thus, it is an operation $E$ could physically perform).
\end{enumerate}

Now, on any particular iteration of a SQKD protocol, Alice will send a state of the form $\alpha\ket{0} + \beta\ket{1}$.  These $\alpha$ and $\beta$ may be chosen randomly if the protocol is a multi-state one, or they may be publicly known if the protocol is a single state one.  After the initial CNOT gate, the joint system becomes $\alpha\ket{0,0}_{TE} + \beta\ket{1,1}_{TE}$.  If $B$ chooses $\R$, the state returning to $E$ is exactly this; $E$ will then apply $U_R$ evolving the joint state to $(\alpha\ket{0} + \beta\ket{1})\otimes\ket{0}_E$ thus creating no detectable disturbance.

On the other hand, if $B$ chooses to $\M$ and $\texttt{Prepare}$, he will detect $\ket{0}$ with probability $|\alpha|^2$ or $\ket{1}$ with probability $|\beta|^2$, the same probability had $E$ chosen to not attack and, so far at least, her attack is not detected.  We may write the resulting state as a density operator:
\[
|\alpha|^2\kb{0}_B\otimes\kb{0}_E + |\beta|^2\kb{1}_B\otimes\kb{1}_E,
\]
where we introduced a $B$ register storing $B$'s measurement result.  Next, $B$ chooses to prepare a fresh qubit (unlike $\MR$, the state he sends may be different from that he observed).  If he sends a $\ket{0}$, the resulting state, again modeling as a density operator due to $B$'s measurement, is:
\[
|\alpha|^2\kb{0}_B\otimes\kb{00}_{TE} + |\beta|^2\kb{1}_B\otimes\kb{01}_{TE},
\]
which becomes, after applying $U_R$:
\begin{equation}
\rho_0 = |\alpha|^2\kb{0}_B\otimes\kb{00}_{TE} + |\beta|^2\kb{1}_B\otimes\kb{02}_{TE}.
\end{equation}
Following the same logic, had $B$ chosen to send $\ket{1}$, we have density operator:
\begin{equation}
\rho_1 = |\alpha|^2\kb{0}_B\otimes\kb{13}_{TE} + |\beta|^2\kb{1}_B\otimes\kb{10}_{TE}.
\end{equation}
The $T$ qubit is passed to $A$.  It is not difficult to see that this attack goes undetected.  However, by measuring her ancilla, if $E$ observes $\ket{2}_E$ she knows for certain that $B$ choose to $\M$ and $\texttt{Prepare}$, that he originally observed $\ket{1}$ and that he sent $\ket{0}$.  If $E$ observes $\ket{3}_E$ she knows that $B$ originally observed $\ket{0}$ and sent $\ket{1}$.  In these cases, $E$ has full information on $B$ and $A$, thus causing a security break.  Of course this attack does not always work.  Indeed, it fails whenever $B$ chooses to send exactly the same state he measured (i.e., he ends up using $\MR$)!  Thus, by increasing $B$'s capabilities, we actually cause a security break.  Since this attack does not work all of the time (some iterations will give $E$ no information) it leads to a partially secure system (i.e., it may be \emph{partially robust} as defined in \cite{SQKD-first} though we will return to this notion of ``robustness'' later in this work).

\openproblem While security is broken when $B$ sends an alternative state than what he observed for key distillation, are there potential advantages to this, perhaps, in better categorizing $E$'s attack?  Additional noise statistics may be gathered which could help security so long as those iterations where $B$ sends the opposite state are never used for the key.

Research has shown that decreasing $A$'s source preparation ability (e.g., single-state protocols) can still lead to secure key distribution systems.  Interestingly it is also possible to decrease her ability to measure.  In \cite{sqkd-limited-measure}, it was shown that $A$ need only measure in the $X$ basis and send three states.  In \cite{sqkd-classical-quantum}, it was shown that $A$ needed only to prepare two states, a $\ket{0}$ and a state $\ket{a}$ where $|\braket{0|a}| \in (0,1)$ while only measuring using a three outcome POVM.  This last paper provided an SQKD protocol that could smoothly transition from classical communication to (semi) quantum communication and proposed a method of measuring the affects this transition has on secure communication rates.  However, their security analysis required a three outcome POVM - while weaker than a two basis measurement, it is not as weak as simply measuring in a single basis as in \cite{sqkd-limited-measure} (though \emph{that} paper required three states).  This leads to a rather interesting open problem:

\openproblem Does there exist an SQKD protocol where $A$ sends only two (non orthogonal) states and only performs a measurement in a single basis?

Clearly, a single state SQKD protocol where $A$ measures only in one basis cannot exist.  This is easy to see: assume such a protocol exists - then, since $A$ sends only a single state, no key material can be transferred on the forward channel, instead it must be transmitted somehow using the reverse or loop channels.  But, since $A$ can only measure in a single basis, and since this basis is public knowledge (due to Kerckhoffs' principle \cite{katz-crypto}, the basis choice should be public knowledge), Eve could simply also measure in this basis any qubit arriving to $A$.  Thus $A$ and $E$'s systems will be fully correlated and no key can be distilled.  Thus, to have any hope of further reducing resource requirements on the quantum user, a two state, one basis protocol is the only possibility.  It is unclear if such a protocol can exist.

One candidate was proposed in \cite{sqkd-classical-quantum}, but it is not clear that it is secure.  While it was shown to be secure against a single Intercept/Resend attack in \cite{sqkd-classical-quantum}, beyond this no proof of security (or insecurity) exists.  Since single-state with two basis measurement protocols exist \cite{SQKD-less-than-4,krawec2014restricted,sqkd-single-state-b92,cl-A}, and since three state, one basis measurement protocols exist \cite{sqkd-limited-measure}, a two-state, one basis measurement protocol would be an interesting development in SQKD research and represent the minimal resource requirements on the part of the quantum user, in a point-to-point SQKD system (there are other models of semi-quantum communication involving third parties which are outside of this question's scope and which we discuss later).

Continuing our discussion on reducing resource requirements, a protocol where $B$ does not need to actually perform a measurement was shown in \cite{no-measure}.  This protocol required the $\M$ and $\Perm$ operations.  Namely, on receipt of a qubit stream from $A$, $B$ will choose to discard a random subset of the qubits and prepare fresh $Z$ basis ones in place of them (these qubits are not first measured).  He then applies the $\Perm$ operation sending these qubits back to $A$.  Another protocol in \cite{no-measure-2} was introduced where $B$ needs only to be able to $\R$ or discard a qubit (he need not measure nor must he prepare a qubit).  This required a third party however.  The protocol consisted of $A$ sending qubits to $B$ who can choose to drop them, or forward them (the $\R$ operator in this case) to the third party server.  This third party was responsible for applying a unitary gate to the qubits received and sending them back to $B$.  This classical user $B$ may again choose to discard or forward them back to the third party who then measures the qubits.

\subsection{Other SQKD Protocols}

While a large portion of research has been in the direction of further reducing resource requirements for an SQKD protocol (either for the fully quantum or the semi quantum user), other novel protocols with interesting theoretical insights have also been developed. One area of research has been in attempting to develop ``authenticated SQKD protocols'' which do not utilize an authenticated channel (instead relying on a pre-shared key).  Protocols of this nature have been proposed in \cite{sqkd-no-auth-1,sqkd-no-auth-2}.  Security in this regime, however, is difficult to define and attacks have been shown in \cite{sqkd-no-auth-atk}.  There has not been a complete information theoretic security analysis for these protocols as of writing this, instead security is generally shown against certain classes of attacks.

Strategically utilizing the two-way quantum channel is another possible research direction to take when designing new SQKD protocols.  In \cite{sqkd-high-noise} a new SQKD protocol was developed where $A$ uses information from both the forward and reverse channels.  This results in a loss of efficiency but results in a drastic increase in protocol noise tolerance - something we will comment on later when discussing security results.

Efficiency is an important consideration to take into account when designing new SQKD protocols.  In \cite{sqkd-eff-0,sqkd-eff} protocols were proposed to improve efficiency by biasing choices to improve their overall efficiency (e.g., by leading to fewer discarded iterations due to incompatible choices), similar to what is done for fully-quantum protocols \cite{QKD-BB84-Modification}.

Other encoding schemes are also worth investigating.  In \cite{sqkd-dephase}, two qubits were used in a form of time-bin encoding.  The encoding was done in a way so as to allow for robust protection against dephasing noise.  A six-qubit encoding scheme was presented in \cite{sqkd-dephase-rotate} (subsequently improved in \cite{sqkd-dephase-rotate-2}) allowing for protection against dephasing and rotation noise.

While the majority of SQKD research is in reducing resource requirements, some work in \cite{sqkd-qw,sqkd-high-dim} has been done in high-dimensional (beyond a small fixed constant number of qubits per signal) SQKD.  Since it is now known that the use of high-dimensional quantum states for fully quantum key distribution provides several benefits, especially in noise tolerance (see \cite{high-dim0,QKD-high-dim-50,high-dim5,high-dim6,high-dim7} for a few references), it is interesting to see that this also translates to the semi-quantum case.

When working with high-dimensional semi-quantum communication, one must define what the classical user's capabilities are when interacting with high-dimensional states.  The natural approach (used also in \cite{sqkd-qw,sqkd-high-dim}) is for $A$, the fully quantum user, to send an arbitrary state $\ket{\psi}$, now living in a $d$-dimensional space (instead of the usual $d=2$).  $B$, when receiving this state, can either choose $\R$ in which case he reflects the entire $d$-dimensional state, or he can choose $\MR$ in which case he performs a measurement in the $d$-dimensional computational basis, namely $\{\ket{0}, \ket{1}, \cdots, \ket{d-1}\}$ and prepares a fresh $d$-dimensional state based on his measurement outcome.  Of course when $d=2$ this agrees completely with Boyer et al.'s original definitions \cite{SQKD-first}; furthermore, if both $A$ and $B$ are restricted to these operations, the protocol is no different, mathematically, from a classical one and, so, this seems the natural way to extend semi-quantum communication to higher dimensions.  In \cite{sqkd-qw}, a high-dimensional SQKD protocol based on the use of quantum walks was presented (here, the states $A$ prepared were results from evolving a quantum walk \cite{QW-intro1,QW-survey}).  In \cite{sqkd-high-dim} a high-dimensional version of BKM07 was presented which was shown to tolerate high levels of noise as the dimension of the quantum state increases (similar to what occurs in the fully-quantum setting \cite{QKD-high-dim-50}).  Future work in developing high-dimensional semi-quantum protocols may prove very interesting in further discovering the differences, and similarities, between the semi-quantum and fully-quantum models of communication.

\section{Security Results}

In this section we discuss research on security aspects of SQKD protocols in the perfect qubit scenario (we leave practical security issues for a later section).  There are two main challenges to performing a security analysis of a semi-quantum protocol.  First is the fact that at least one party (potentially both as discussed above) is limited in some nature and, therefore, users cannot make certain measurements on the noise in the quantum channel.  For instance, they cannot measure the $X$ basis noise in the forward channel (from $A$ to $B$).  Second is the fact that $E$ has two opportunities to interact with the qubit in flight - first when it travels to $B$ and second on its return.  Indeed, as shown in \cite{sqkd-eavesdropping}, attacking twice can allow an adversary greater information gain than simply attacking one channel, at least for some SQKD protocols.  In this section we review general techniques for arguing about the security of SQKD protocols.

\subsection{Robustness}

The first notion of security for a semi-quantum protocol was \emph{robustness}.  This was a term introduced by Boyer et al., in the original SQKD paper \cite{SQKD-first} and states that an SQKD protocol is \emph{robust} if any attack which causes an adversary to learn non-zero information on the raw key of the protocol must necessarily induce a detectable disturbance in the quantum channel.  That is, the adversary cannot get any information without risking detection.  The notion of \emph{partial robustness} was introduced in that same paper which weakens the definition to allowing the adversary to gain some information without being detected, but any attack which gains full information on the raw key must induce a detectable disturbance.

\begin{figure}
  \centering
  \includegraphics[width=225pt]{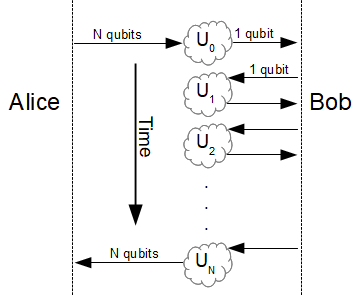}
\caption{Showing the attack scenario considered in Boyer et al.'s proof of robustness \cite{sqkd-second}.  Each $U_i$ is a unitary probe acting on all $N$ qubits and $E$'s quantum ancilla.}\label{fig:sqkd-robust-atk-1}
\end{figure}

To prove an SQKD protocol robust, one must show that for any attack, if $E$'s ancilla is somehow correlated with $A$ or $B$'s raw key bit register, then this attack cannot go undetected with unit probability.  In general, there are two main methods of proving a protocol robust.  The first, introduced in \cite{sqkd-second} involves the following:

\begin{enumerate}
  \item First, assume that $E$ is able to capture all $N$ qubits leaving $A$'s lab in bulk.  Before forwarding them to $B$, $E$ applies a unitary probe $U_0$ acting jointly on all qubits and $E$'s private ancilla.
  \item Following this probe, $E$ forwards the first qubit to $B$.  After $B$'s operation, the qubit returns to $A$ however $E$ captures it again.  At this point, the adversary once again holds all $N$ qubits and applies a new unitary probe $U_1$ which, as with $U_0$, acts jointly on all qubits and $E$'s private ancilla (the same ancilla throughout).
  \item $E$ then repeats the above, sending the next qubit to $B$, capturing it on its return, and probing it with a new unitary operator $U_i$.
  \item Once all $N$ qubits have gone through this process, $E$ returns the $N$ qubits to $A$ who completes the protocol.
\end{enumerate}
This process is depicted in Figure \ref{fig:sqkd-robust-atk-1}.  It is obviously a very strong attack model allowing $E$ to capture these qubits in bulk.

The second method of showing robustness was introduced first in \cite{SQKD-less-than-4} by Zou et al., and makes the assumption that $A$ only sends a subsequent qubit to $B$ \emph{after} receiving the previous one back from him.  This assumption restricts $E$ from storing all $N$ qubits simultaneously, however the probe $E$ uses need not be the same for every iteration.  This assumption generally simplifies the security proof and allows for an inductive style argument to prove robustness.  Indeed, consider the BKM07 protocol - an inductive style of robustness proof could proceed along these lines:  Consider the first iteration of the protocol.  Then, $A$ sends a qubit state of the form $\ket{i}_T$ for $i\in\{0,1,+,-\}$ (we use ``$T$'' to represent the transit space - i.e., the two-dimensional Hilbert space modeling the qubit in transit between parties).  Let $U_F^{(1)}$ be $E$'s first probe in the forward direction.  Since this is the first iteration, $E$'s ancilla is cleared to some default pure state $\ket{\chi}_E$ which we may assume is known to $E$ (note, the state is pure to $E$'s advantage).  The action of this first probe on basis states may be written as:
\begin{align*}
U_F^{(1)}\ket{0}_T\otimes\ket{\chi}_E &= \ket{0,e_0}_{TE} + \ket{1,e_1}_{TE}\\
U_F^{(1)}\ket{1}_T\otimes\ket{\chi}_E &= \ket{0,e_2}_{TE} + \ket{1,e_3}_{TE}\\
\end{align*}
where the $\ket{e_i}$ are arbitrary states in $E$'s private ancilla (these are not assumed to be normalized or orthogonal).  However, with non-zero probability this iteration will be used for error detection.  Thus, to avoid detection, $E$ must set $\ket{e_1} \equiv \ket{e_2} \equiv \mathbf{0}$, that is, both must be the zero vector.  When the qubit returns (recall, this second model of robustness from \cite{SQKD-less-than-4} assumes $A$ will not send another qubit until this one returns - thus $E$ is forced to probe the qubit immediately on return from $B$), $E$ applies a second probe $U_R^{(1)}$ whose action we may write as:
\begin{align*}
U_R^{(1)}\ket{0,e_0}_{TE} &= \ket{0, f_0}_{TE} + \ket{1,f_1}_{TE}\\
U_R^{(1)}\ket{1,e_3}_{TE} &= \ket{0, f_2}_{TE} + \ket{1,f_3}_{TE},
\end{align*}
where, like the $\ket{e_i}$ states, the $\ket{f_j}$ are arbitrary states in $E$'s ancilla.  Note that $U_R^{(1)}$'s action on states not of the form $\ket{0,e_0}$ and $\ket{1,e_3}$ is irrelevant as they never appear.  As before, this iteration may be used for error detection with non zero probability. Thus to avoid detection, namely to avoid inducing any $Z$ basis noise in the reverse channel, $E$ must set $\ket{f_1} \equiv \ket{f_2} \equiv \mathbf{0}$.  Now, with non-zero probability $A$ might have sent $\ket{+}$ and $B$ may have chosen $\R$.  In this case, the state returning to $A$ is found to be:
\begin{align}
U_R^{(1)}U_F^{(1)}\ket{+,\chi}_{TE} &= U_R^{(1)}\left(\frac{1}{\sqrt{2}}\ket{0,e_0} + \frac{1}{\sqrt{2}}\ket{1,e_3}\right)\notag\\
&= \frac{1}{\sqrt{2}}\ket{0,f_0} + \frac{1}{\sqrt{2}}\ket{1,f_3}\notag\\
&= \frac{1}{2}\ket{+}(\ket{f_0} + \ket{f_3}) + \frac{1}{2}\ket{-}(\ket{f_0}-\ket{f_3}),
\end{align}
where the last equality arises from changing the transit space from the $Z$ to the $X$ basis.  At this point, it is clear that to avoid detection, it must hold that $\ket{f_0} = \ket{f_3}$.  Thus, $E$'s ancilla after this first iteration is completely independent of the transit space and $A$ and $B$'s measurements.  Through induction, one sees that this holds for each subsequent iteration for probes $U_F^{(i)}$ and $U_R^{(i)}$ (though note that the initial state $\ket{\chi}_E$ is potentially different each iteration, but remains independent of $A$ and $B$'s state).

Proving robustness in the first model introduced by Boyer et al., in \cite{SQKD-first} is more involved.  It is an interesting open problem, however, to know whether or not this iterative attack as assumed in the Boyer model of robustness gives $E$ any advantage.

\subsection{Information Theoretic Analysis}

The notion of robustness gives a good security guarantee in that any adversary who attempts to learn something about the raw key risks being detected.  Beyond this, it is often useful, however, to know exactly \emph{how much} an adversary could have learned given a certain amount of detectable noise.  As current quantum communication systems are not perfect there will always be natural noise that cannot be avoided.  As we assume all-powerful adversaries, we must actually assume, therefore, the worst case that the adversary replaces the noisy quantum channel with a perfect one.  She then hides her attack within the expected natural noise.  Thus, as is typical with standard QKD research \cite{qkd-survey,qkd-survey2,qkd-survey3,qkd-survey4}, we must assume that any detectable noise is the result of an adversary.  The question then is how much information can an adversary gain?  And, furthermore, how much noise is ``too much.''  Similar questions also involve the protocol's efficiency as determined by its key-rate.

Recall that a (S)QKD protocol is considered secure if, after privacy amplification, Equation \ref{eq:qkd-secure} holds.  If we consider collective attacks \cite{qkd-survey} (i.i.d. attacks where $E$ is free to postpone measurement of her ancilla until any future point in time and, furthermore, is free to perform a joint coherent measurement on her ancilla at that point to attempt to extract maximal information), let $N$ be the size of the raw key \emph{before} error correction and privacy amplification are run.  Let $\ell$ be the size of the secret key satisfying Equation \ref{eq:qkd-secure}; then it was shown in \cite{QKD-renner-keyrate} that, in the asymptotic limit it holds that:
\begin{equation}\label{eq:key-rate-entropy}
\lim_{N\rightarrow \infty}\frac{\ell}{N} = \inf\left[S(A|E) - H(A|B)\right],
\end{equation}
where the infimum is over all collective attacks which induce the observed noise statistics (e.g., error rates).  Here, $S(A|E)$ is the von Neumann entropy of $A$'s raw-key bit register conditioned on $E$'s quantum memory system while $H(A|B)$ is the classical Shannon entropy of $A$'s raw-key bit register conditioned on $B$'s.  This equation is very intuitive: it states that the key-rate increases when $E$ has a lot of uncertainty (measured by $S(A|E)$) and $B$ has little uncertainty (measured by $H(A|B)$).  The goal of a (S)QKD security proof in this manner is to determine a lower-bound on $r$, given only the observed noise statistics.  That is, one cannot compute $S(A|E)$ with certainty since we do not know exactly what attack $E$ used - however, one can attempt to lower-bound $E$'s uncertainty based on the noise assuming she chose an optimal attack which induces that observed noise.  Note that an alternative, and equivalent, version of Equation \ref{eq:key-rate-entropy}, derived in \cite{QKD-Winter-Keyrate}, is:
\begin{equation}
\lim_{N\rightarrow \infty}\frac{\ell}{N} = \inf\left[I(A:B) - I(A:E)\right],
\end{equation}
where $I(A:E)$ is the quantum mutual information and $I(A:B)$ is the classical mutual information.  That these two versions of the key-rate expression are the same follows immediately from the definition of mutual information and the fact that the systems under consideration, namely $A$ and $B$'s classical raw-key registers and $E$'s quantum register, are classical-classical-quantum states.

The question, then, becomes: given certain observed channel statistics (e.g., noise rates in the quantum channel(s)), what is a protocol's key-rate?

\subheading{Individual Attacks}
The first papers to attempt to answer this question were \cite{cl-A} and \cite{info-disturbance} which both considered information gain as functions of observed noise assuming \emph{individual attacks}, namely attacks whereby the adversary attacks each qubit identically and, furthermore, is forced to measure her ancilla immediately thus leading to a classical memory.

\begin{figure}
  \centering
  \includegraphics[width=250pt]{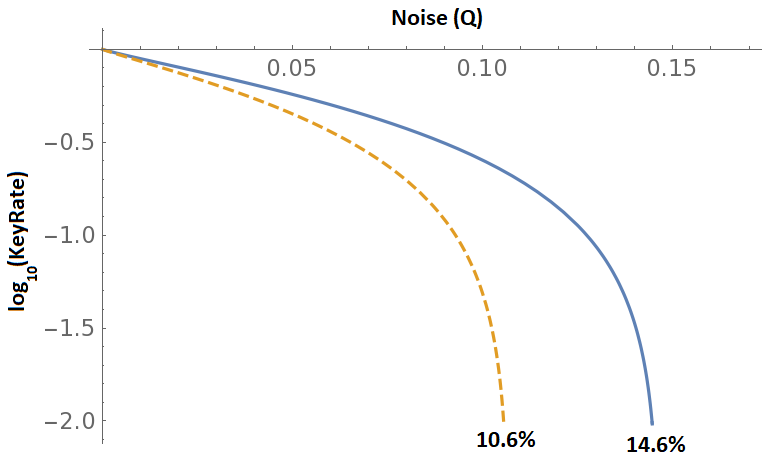}
\caption{Showing the key-rate of the two-state protocol introduced in \cite{cl-A} assuming individual attacks.  For the Solid Line, we consider $Q_X = Q$; for the Dashed Line, we consider $Q_X = 2Q(1-Q)$.}\label{fig:cl-a-keyrate}
\end{figure}

In \cite{cl-A}, the authors introduced a new SQKD protocol which we discussed in an earlier section.  Security for their protocol was proven in terms of an individual attack on the reverse channel (an argument was made that attacking the forward channel for their specific protocol under the assumption of individual attacks did not provide her with any additional information).  Under this attack, they derived the following expression for the mutual information held between $A$ and $E$, namely:
\begin{equation}
I(A:E) = 1 - h\left(\frac{1 + x}{2}\right),
\end{equation}
where:
\begin{equation}
x = 2\sqrt{Q_X(1-Q_X)},
\end{equation}
and $Q_X$ is the observed $X$ basis error rate in the channel whenever $B$ chooses $\R$.  Of course $I(A:B)$ is simply $1-h(Q)$ where $h(\cdot)$ is the binary Shannon entropy function and $Q$ is the probability of a $Z$ basis error in the reverse channel (note that the reverse channel is used to carry key material for this particular protocol).  A graph of the resulting key rate $r = I(A:B) - I(A:E)$ is shown in Figure \ref{fig:cl-a-keyrate}.  Interestingly, the noise tolerance of this protocol in this attack model is $14.6\%$ (when $Q_X = Q$) which is exactly that which BB84 can tolerate against individual attacks \cite{BB84-ind1}.  This connection in noise tolerance between semi-quantum and fully-quantum key distribution is something we will comment on again later when looking at stronger security models and shows that, even though semi-quantum protocols are limited in their quantum capabilities, \emph{they hold similar security properties to that of fully quantum protocols}, at least in ideal qubit channels (practical issues surrounding semi-quantum cryptography remain a large area of open research which we address later in this paper).  It would be interesting to investigate SQKD protocols in other attack models where an adversary is limited in their quantum abilities and compare to the fully-quantum counterpart.

\openproblem It has been demonstrated that the noise tolerance of SQKD protocols are comparable, or equal, to that of fully quantum protocols in the individual attack scenario and, as we discuss later, against stronger collective attacks.  Does this relation hold for security models that may be weaker than individual attacks?  For instance, intercept-resend attacks (where $E$ must measure in a particular basis and forward a result - i.e., she cannot probe the qubit sent any other way).

\begin{figure}
  \centering
  \includegraphics[width=250pt]{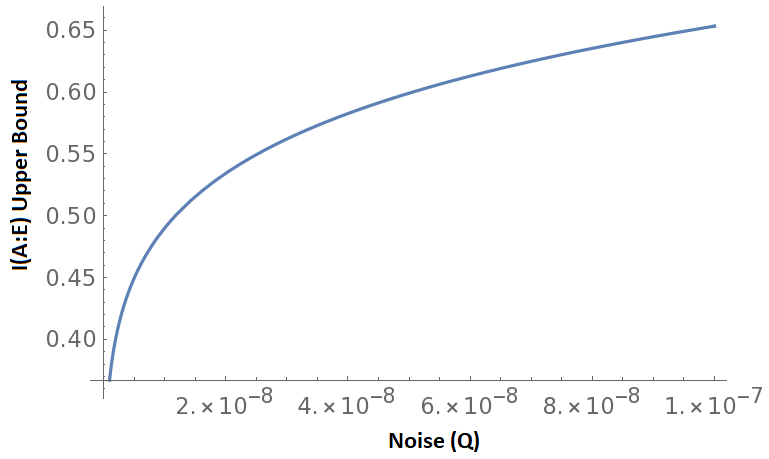}
\caption{Showing the \emph{upper bound} on the mutual information held between $A$ and $E$ as derived in \cite{info-disturbance} for Zou's single-state protocol introduced originally in \cite{SQKD-less-than-4}.  This assumes $E$ is restricted to individual attacks. New work has since determined more optimistic bounds, but this original result remains important as one of the first to derive a relation between disturbance and information gain for an SQKD protocol.}\label{fig:info-dist-LB}
\end{figure}

In \cite{info-disturbance}, an analysis of Zou's single state protocol (introduced in \cite{SQKD-less-than-4}) was performed also assuming individual attacks.  There, the mutual information between $A$ and $E$ was found to be:
\begin{equation}
I(A:E) \le 2\sqrt{Q_X + 6Q^{1/4}}.
\end{equation}
This is the first upper-bound on the mutual information for the single-state SQKD protocol, at least for individual attacks.  It also hints at robustness as, when $Q_X=Q=0$, then $I(A:E) = 0$; that is, when there is no noise, $E$ cannot extract information.  Of course, robustness should not assume that $E$ measures her ancilla which is an assumption made when handling individual attacks.  Note that this result is only an upper-bound and, as shown in Figure \ref{fig:info-dist-LB}, is very pessimistic as $E$'s information gain is potentially very large even for small disturbances.  More optimistic bounds for this protocol have since been shown also assuming a stronger attack model as we discuss later.  However, the paper \cite{info-disturbance} was one of the first to actually derive a connection between noise and information gain for a semi-quantum protocol and remains an important work.

\subheading{Restricted Attacks}

One of the challenges with proving security of an SQKD protocol, either in terms of robustness, or some other security model, is that the adversary is allowed two opportunities to attack the qubit as it travels.  However, there are currently two techniques for reducing this complexity.

We consider, now, \emph{collective attacks} where $E$ attacks the forward channel with an operator $U_F$ and the reverse with an operator $U_R$.  Both of these are unitary and act on the two dimensional qubit space and $E$'s private ancilla (of arbitrary dimension).  This can be used to prove robustness in Zou's model \cite{SQKD-less-than-4} or for an information theoretic analysis of noise tolerance against collective attacks which we discuss next.

First, it was shown in \cite{krawec2014restricted} that for any \emph{single-state} SQKD protocol using only the $\MR$ and $\R$ operations, the forward attack operator does not need to entangle the traveling qubit with $E$'s quantum ancilla.  Instead, it is sufficient to bias the state's probability amplitudes.  However, beyond this, the attack does not provide $E$ with any additional information.  Note this result is not true if $B$ is allowed to use an operation beyond $\MR$ or $\R$.  Of course it is also not true if $B$ is more powerful than semi-quantum.

More formally, as proven in \cite{krawec2014restricted}, to prove security against collective attacks, or robustness in Zou et al.'s model, it suffices to consider an attack whereby $E$ sends to $B$ a qubit state:
\begin{equation}\label{eq:bias-atk}
\ket{\psi} = \sqrt{\frac{1}{2}+b}\ket{0} + \sqrt{\frac{1}{2}-b}\ket{1},
\end{equation}
for some real bias parameter $b \in [-1/2, 1/2]$.  Any arbitrary collective attack of the form $(U_F, U_R)$ can be ``reduced'' to this \emph{restricted collective attack} without loss of advantage to an all-powerful adversary.  This allows for simplified security analyses as one must only consider the forward channel bias and not any entanglement with an adversarial system.  In fact, one may even enforce a symmetry to $E$'s attack in that, if $A$ was supposed to send a $\ket{+}$ state, $B$ can enforce that $b = 0$ and abort otherwise (such a symmetry assumption is often made in S(QKD) security proofs).  Regardless, however, it is clear that this value $b$ is an observable parameter that may be used in any security proof.

This restricted attack definition was recently extended to arbitrary \emph{multi-state} SQKD protocols in \cite{krawec-entropic}.  For multi-state protocols, $E$ need only apply a restricted forward attack operator $\mathcal{F}$ which acts as follows:
\begin{align*}
\mathcal{F}\ket{0}_T\otimes\ket{\chi}_E &= q_0\ket{0,0}_{TE} + \sqrt{1-q_0^2}\ket{1,e}_{TE}\\
\mathcal{F}\ket{1}_T\otimes\ket{\chi}_E &= q_2\ket{1,f}_{TE} + \sqrt{1-q_2^2}\ket{1,0}_{TE}.
\end{align*}
where, $q_i$ are positive real numbers no greater than one and, furthermore, one may assume that:
\begin{align*}
\ket{e}_E &= \eta_0\ket{0}_E + \sqrt{1-|\eta_0|^2}\ket{1}_E\\
\ket{f}_E &= \eta_1\ket{0}_E + \sqrt{1-|\eta_1|^2}\ket{1}_E
\end{align*}
where $\eta_0$ and $\eta_1$ are complex numbers such that $|\eta_i| \le 1$.  Note that, unlike the single-state case, for a multi-state protocol one must consider $E$ entangling the qubit with her quantum ancilla in the forward channel.  However, the dimension of $E$'s memory need only be two dimensional.  Furthermore, rather interestingly, the state of her ancilla is essentially a ``right'' or ``wrong'' state - namely it is $\ket{0}_E$ when $\mathcal{F}$ does not flip the input while it is one of the $\ket{e}$ or $\ket{f}$ otherwise.  Such simplified attacks can help to perform a security analysis of \emph{any} SQKD protocol which relies on operations $\MR$ and $\R$.  For single-state protocols, one should use the restricted bias-only version \cite{krawec2014restricted}; for others, one must use the alternative definition for multi state protocols from \cite{krawec-entropic}.

\openproblem Do equivalent restricted attacks exist for protocols where $B$ uses $\Perm$?  Proofs of equivalency from \cite{krawec2014restricted,krawec-entropic} normally employ the following strategy: Fix an attack against the protocol.  Work out the density operator describing the protocol when the particle(s) return to Eve for the second time.  Next, show that, if a restricted attack were used, the returning state can be ``fixed'' via a unitary operator so that it is equal to the general case.  If this is possible, there is no advantage to $E$ using a full attack, she might as well use the simplified attack.  Note that one must also be careful to ensure that $A$ and $B$ cannot tell the difference (e.g., the two attacks should induce the same observable statistics in both cases).  Can a definition of restricted attack for the $\Perm$, or other semi-quantum, operations be defined and proven?

\subheading{Key Rate Computations}

Moving beyond robustness, it is important to understand how a protocol behaves when faced with noise.  In detail, one wishes to derive an information theoretic bound on the key-rate of a protocol (see Equation \ref{eq:key-rate-entropy}) as a function only of observable parameters.  This allows for the better understanding of a protocol's performance (e.g., its noise tolerance) and also allows us to better compare semi-quantum protocols with fully-quantum ones.  Since one of the main theoretical goals of the semi-quantum model of cryptography is to better map out the ``gap'' between fully quantum and partially quantum protocols, having a rigorous way to gauge relative performance of two protocols is vital.  Key-rate under certain noise conditions makes for an excellent measure to compare.

The first information theoretic analysis of an SQKD protocol was in 2015 in \cite{krawec2015security} with several other protocols analyzed since then.  Assuming collective attacks in the asymptotic scenario, doing so ultimately requires bounding the entropy term $S(A|E)$, where $\rho_{AE}$ is a density operator describing a single iteration of the protocol conditioning on that iteration being used for raw-key distillation.  In general, there seem to be three main methods currently for deriving key-rate computations for SQKD protocols:
\begin{enumerate}
  \item Compute a lower bound on $S(A|E)$ based on strong subadditivity.
  \item For single state-protocols, use biased-restricted attacks to argue that when the bias is $0$ (see Equation \ref{eq:bias-atk}), the SQKD protocol is equivalent to a known one-way protocol for which the entropy is known; next, argue using the continuity of von Neumann entropy \cite{QC-alicki2004continuity,QC-fannes-audenaert-inequality,QC-winter2016tight}, that as the bias changes, the entropy cannot differ ``too much.''  Thus the key-rate cannot decrease ``too much'' based on $b$.
  \item Reduce the protocol to an equivalent one-way entanglement based protocol (shown in \cite{krawec-entropic,sqkd-high-dim} to be possible at least for \emph{some} SQKD protocols) and use entropic uncertainty relations \cite{QKD-uncertainty,uncertainty-survey1,uncertainty-survey-2,uncertainty-survey-3}.
\end{enumerate}

\subheading{First Key-Rate Proof Method}
Perhaps the most generally applicable approach is to compute $S(A|E)$ directly and this seems to be the approach used for the majority of SQKD key-rate computations.  First used in \cite{krawec2015security} but improved in \cite{QKD-Tom-Krawec-Arbitrary}, one must begin by writing out a density operator description of a single iteration of the protocol assuming that iteration is used to distill a raw key bit.  Namely, one must condition on events leading to a raw-key iteration.  Taking into account also $E$'s attack (one may take advantage, here, of the restricted attack results for single-state \cite{krawec2014restricted} or multi-state \cite{krawec-entropic} protocols as discussed earlier) this results in a classical-classical-quantum state (ccq-state) of the form:
\begin{equation}\label{eq:ccq}
\rho_{ABE} = \frac{1}{N}\sum_{i,j\in\{0,1\}}\kb{i,j}_{AB}\otimes\rho_E^{(i,j)},
\end{equation}
where $\rho_E^{(i,j)}$ is a density operator modeling $E$'s attack in the event $A$'s raw key bit happens to be $i$ and $B$'s raw key bit is $j$ and where $N$ is a normalization term.

The operators $\rho_E^{(i,j)}$ can always be written as a sum of the form:
\[
\rho_E^{(i,j)} = \sum_k \kb{E^{(i,j)}_k},
\]
where $\ket{E_k^{(i,j)}}$ are vectors (possibly sub normalized if the above sum has more than one element) in $E$'s ancilla.  The fact that one may write the operators $\rho_E^{(i,j)}$ in this form is a basic fact of linear algebra.  However, the exact structure of these operators usually is found through the derivation of $\rho_{ABE}$ and so, generally, no additional work is needed to decompose the operators in this structure (after tracing the protocol to derive $\rho_{ABE}$).

In \cite{QKD-Tom-Krawec-Arbitrary} a general theorem was derived allowing one to compute the conditional entropy of a state shown in Equation \ref{eq:ccq}:
\begin{theorem}\label{thm:entropy}
(From \cite{QKD-Tom-Krawec-Arbitrary}): Let $\rho_{AE}$ be a state of the form:
\begin{equation}\label{eq:cq-thm}
\rho_{AE} = \frac{1}{N}\kb{0}_A\otimes\left(\sum_{k=0}^{M}\kb{E_k^{(0)}}\right) + \frac{1}{N}\kb{1}_A\otimes\left(\sum_{k=0}^{M}\kb{E_k^{(1)}}\right)
\end{equation}
and denote by $N_k^i$ to mean $N_k^i = \bk{E_k^{(i)}}$.  Then it holds that:
\begin{equation}
S(A|E) \ge \sum_{k=1}^{M}\left(\frac{N_k^0+N_k^1}{N}\right)S_k,
\end{equation}
where:
\begin{equation}
S_k = \left\{\begin{array}{ll}
h\left(\frac{N_k^0}{N_k^0+N_k^1}\right) - h(\lambda_k) & \text{ if both $N_k^0 > 0$ and $N_k^1 > 0$}\\
0 & \text{ otherwise}
\end{array}\right.
\end{equation}
and finally:
\begin{equation}
\lambda_k = \frac{1}{2}\left(1 + \frac{\sqrt{\left(N_k^0-N_k^1\right)^2 + 4Re^2\braket{E_k^{(0)}|E_k^{(1)}}}}{N_k^0+N_k^1}\right).
\end{equation}
\end{theorem}

Note that this theorem is very general and can be applied to any cq-state - either one produced by a SQKD protocol, or one produced by some other protocol, quantum or semi-quantum.  It states that one may compute the conditional entropy simply by knowing (or bounding) the inner products of states $\bk{E_k^{(i)}}$ (in an SQKD protocol these are typically found by looking at the $Z$ basis noise in the quantum channel) and the overlap between $\ket{E_k^{(0)}}$ and $\ket{E_k^{(1)}}$ (which in an SQKD protocol can usually be bounded by looking at the $X$ basis noise in the case when $B$ chooses $\R$).

We make three comments on the above theorem.  First, one may always write a density operator $\rho_{AE}$ in the form shown in Equation \ref{eq:cq-thm}.  This is due to the fact that the theorem allows some $\ket{E_k^{(j)}}$ to be zero vectors and so the total number of terms in the $0$ and $1$ case may both be $M$.

Secondly, the ordering of the terms appearing in the $E$ portion of the density operators in Equation \ref{eq:cq-thm} is irrelevant.  Indeed, one may apply any permutation $\pi:\{1,\cdots,M\}\rightarrow\{1,\cdots,M\}$ and consider the (equivalent) density operator:
\[
\rho_{AE} = \frac{1}{N}\kb{0}_A\otimes\left(\sum_{k=0}^{M}\kb{E_k^{(0)}}\right) + \frac{1}{N}\kb{1}_A\otimes\left(\sum_{k=0}^{M}\kb{E_{\pi(k)}^{(1)}}\right).
\]
Theorem 1 will provide a lower-bound on $S(A|E)$ for any such ordering, even though, now, one considered inner-products of the form $Re\braket{E_{k}^{(0)}|E_{\pi(k)}^{(1)}}$.  For all permutations, all lower-bounds are correct bounds on the entropy in the state $\rho_{AE}$.  Thus, when applying this theorem to key-rate computations, one must arrange the terms strategically to get the most optimistic lower-bound.  In general, the ``rule of thumb'' appearing in most SQKD papers using this result is to arrange vectors so that $\ket{E_k^{(0)}}$ and $\ket{E_k^{(1)}}$ have both similar weights (i.e., $N_k^0$ is close to or equal to $N_k^1$) and appear in $E$'s system when similar events occur (e.g., when there is no error in the quantum channel or when there is a double-error).  Of course, this is just a guideline - when working with this theorem, it is important to keep in mind that an alternative arrangement of the terms may lead to more optimistic results (but all orderings lead to technically correct lower-bounds on $S(A|E)$).

Third, and finally, one may actually get a more optimistic bound on $S(A|E)$ by defining $\lambda_k$ as:
\begin{equation}
\lambda_k = \frac{1}{2}\left(1 + \frac{\sqrt{\left(N_k^0-N_k^1\right)^2 + 4|\braket{E_k^{(0)}|E_k^{(1)}}|^2}}{N_k^0+N_k^1}\right).
\end{equation}
That is, instead of using only the real part of $\braket{E_k^{(0)}|E_k^{(1)}}$, one should use both real and imaginary to get a tighter bound.  This fact is easily seen from the proof of Theorem \ref{thm:entropy} from \cite{QKD-Tom-Krawec-Arbitrary}.  Though, in key-rate proofs, it is often easier to determine a bound on only the real part, thus the original statement, using only the real part, has, so far, been more useful.

To demonstrate its application, we consider the original Boyer et al., protocol \cite{SQKD-first}, BKM07.  The following proof is from \cite{QKD-Tom-Krawec-Arbitrary}, we highlight the main details here.  Consider a particular collective attack as a pair of unitary operators $U_F$ applied in the forward channel and $U_R$ applied in the reverse.  Since we are considering collective attacks, we may assume $E$'s ancilla is initially cleared to some default state $\ket{\chi}_E$.  Then, without loss of generality, we may write the action of $E$'s attack operators as follows:
\begin{align}
U_F\ket{0,\chi}_{TE} &= \ket{0,e_0} + \ket{1,e_1}\label{eq:unitary-atk}\\
U_F\ket{1,\chi}_{TE} &= \ket{0,e_2} + \ket{1,e_3}\notag\\\notag\\
U_R\ket{i,e_j}_{TE} &= \ket{0,e_{i,j}^0} + \ket{1,e_{i,j}^1}.\notag
\end{align}
where the various $\ket{e_i}$ and $\ket{e_{i,j}^k}$ states are arbitrary states which are not necessarily normalized nor orthogonal in $E$'s ancilla.

Now, one must construct the ccq-state $\rho_{ABE}$.  Full details can be found in \cite{krawec2015security,QKD-Tom-Krawec-Arbitrary}, however, tracing the protocol's execution, including $E$'s attack, and conditioning on the iteration being used for key-distillation (thus, one need only consider $A$ sending a $Z$ basis state, $B$ choosing $\MR$ and $A$ measuring again in the $Z$ basis), one finds the following operator:
\begin{align}
\rho_{ABE} &= \frac{1}{2}\kb{00}_{AB}\otimes\left(\kb{e_{0,0}^0} + \kb{e_{0,2}^0}\right)\\
&+\frac{1}{2}\kb{11}_{AB}\otimes\left(\kb{e_{1,3}^1} + \kb{e_{1,1}^1}\right)\notag\\
&+\frac{1}{2}\kb{01}_{AB}\otimes\left(\kb{e_{1,3}^0} + \kb{e_{1,1}^0}\right)\notag\\
&+\frac{1}{2}\kb{10}_{AB}\otimes\left(\kb{e_{0,0}^1} + \kb{e_{0,2}^1}\right).\notag
\end{align}

Tracing out $B$ yields:
\begin{align}
\rho_{AE} &= \frac{1}{2}\kb{0}_A \otimes\left(\kb{e_{0,0}^0} + \kb{e_{0,2}^0} + \kb{e_{1,3}^0} + \kb{e_{1,1}^0}\right)\\
&+\frac{1}{2}\kb{1}_{A}\otimes\left(\kb{e_{1,3}^1} + \kb{e_{1,1}^1} + \kb{e_{0,0}^1} + \kb{e_{0,2}^1}\right).\notag
\end{align}

Note that the structure of this state is already in the form of Equation \ref{eq:cq-thm} needed to apply Theorem \ref{thm:entropy}.  The states have been paired according to the general rule as mentioned earlier; indeed, note that $\bk{e_{0,0}^0}$ and $\bk{e_{1,3}^1}$ are the highest weighted vectors since they appear when there is no $Z$ basis noise in the forward \emph{and} reverse channel.  The inner products $\bk{e_{i,j}^k}$ (needed to compute the resulting lower bound from the theorem) can all be computed by observing the $Z$ basis noise in the channel.  If $Q$ is the observed $Z$ basis noise in the forward and reverse channel, one finds \cite{QKD-Tom-Krawec-Arbitrary}:
\begin{align*}
&\bk{e_{0,0}^0} = \bk{e_{1,3}^1} = (1-Q)^2\\
&\bk{e_{0,2}^0} = \bk{e_{1,1}^1} = Q(1-Q)\\
&\bk{e_{0,0}^1} = \bk{e_{1,3}^0} = Q(1-Q)\\
&\bk{e_{1,1}^0} = \bk{e_{0,2}^1} = Q^2.
\end{align*}
The above are all found simply by tracing the evolution of the qubit and using Equation \ref{eq:unitary-atk}.  To finish the computation, one requires also bounds on the inner-products $E_1 = Re\braket{e_{0,0}^0|e_{1,3}^1}$, $E_2 = Re\braket{e_{1,1}^1|e_{0,2}^0}$, $E_3 = Re\braket{e_{0,0}^1|e_{1,3}^0}$ and $E_4 = Re\braket{e_{1,1}^0|e_{0,2}^1}$.  These are more involved - for complete details see \cite{QKD-Tom-Krawec-Arbitrary}.  However if one assumes a symmetric channel, the final entropy expression simplifies to:
\begin{align}
S(A|E) &\ge (1-Q)^2[1-h(\lambda_1)] + Q(1-Q)[1-h(\lambda_2)]\label{eq:bkm07-keyrate}\\
&+ Q(1-Q)[1-h(\lambda_3)] + Q^2[1-h(\lambda_4)],\notag
\end{align}
where:
\begin{align*}
\lambda_1 = \frac{1}{2}\left(1 + \frac{|E_1|}{(1-Q)^2}\right) && &\lambda_4 = \frac{1}{2}\left(1 + \frac{|E_4|}{Q^2}\right)\\
\lambda_2 = \frac{1}{2}\left(1 + \frac{|E_2|}{Q(1-Q)}\right) && &\lambda_3 = \frac{1}{2}\left(1 + \frac{|E_3|}{Q(1-Q)}\right).
\end{align*}
Finally, it can be shown that $E_1 = 1 - 2Q_X - E_2-E_3-E_4$, where $Q_X$ is the observed $X$ basis noise whenever $B$ chooses $\R$ and $A$ measures in the $X$ basis (having sent an $X$ basis state initially).  To compute $S(A|E)$, therefore, one must minimize over all $E_2$, $E_3$, and $E_4$ subject to the constraints $|E_2|,|E_3| \le Q(1-Q)$ and $|E_4| \le Q^2$ (these bounds were derived from the Cauchy-Schwarz inequality).  For complete details on this proof method, the reader is referred to \cite{QKD-Tom-Krawec-Arbitrary}.  Computing $H(A|B)$, necessary to finish the key-rate bound, is trivial given the observed error rates in the \emph{raw key} (in this case, it is $H(A|B) = h(Q)$).

\begin{figure}
  \centering
  \includegraphics[width=225pt]{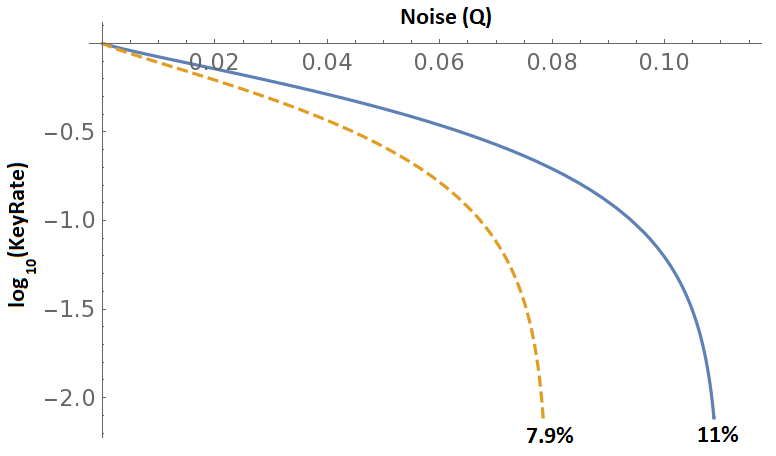}
\caption{Showing an evaluation of the key-rate of the BKM07 protocol as derived in \cite{QKD-Tom-Krawec-Arbitrary}, namely Equation \ref{eq:bkm07-keyrate}.  We consider both dependent (Solid Line, when $Q_X = Q$) and independent (Dashed Line, when $Q_X = 2Q(1-Q)$) channels - see text for explanation.}\label{fig:sqkd-keyrate}
\end{figure}

A plot of the resulting key-rate is shown in Figure \ref{fig:sqkd-keyrate}.  Generally, when evaluating key-rates for protocols relying on a two-way quantum channel (fully quantum or otherwise), one often considers \emph{independent channels} and \emph{dependent channels} \cite{QKD-TwoWaySecure,QKD-Tom-Krawec-Arbitrary}.  For the first, it is assumed that the observed $X$ basis noise in the entire joint forward-reverse channel (when $B$ chooses $\R$) is $2Q(1-Q)$; for the dependent channel, the observed $X$ basis noise is simply $Q$, the error in each channel individually.  Note that certain fiber channels can exhibit this dependent case \cite{QKD-TwoWaySecure,lucamarini2014quantum}.  Of course, these two assumptions are not necessary for security - instead they are just used to evaluate the key-rate and determine noise tolerances.  Since these are commonly used, they also provide good comparison cases.

Evaluating the key-rate expression (Figure \ref{fig:sqkd-keyrate}), one notes that the noise tolerance for a dependent channel is $11\%$ exactly that of BB84 \cite{QKD-BB84,QKD-BB84-rate1,QKD-renner-keyrate}.  For an independent channel, where the $X$ basis noise is roughly twice as high, the noise tolerance drops to $7.9\%$.  However, for a similar $X$ basis noise, this is also the noise tolerance of BB84.  In fact, the key-rate equation shown in Equation \ref{eq:bkm07-keyrate} numerically agrees with BB84 on these two channels.  \emph{This demonstrates that the semi-quantum model, at least from a theoretical perspective, can attain just as high noise tolerances and similar security properties to that of fully quantum protocols!}

In general, Theorem \ref{thm:entropy} allows one to derive a bound on the key-rate expression allowing for fine-grained control over the result through the use of numerous statistics.  This method has been used, now, for several other SQKD protocols beyond BKM07.  Compared with the other methods for proving SQKD security, this has, so far, given the most optimistic results.  However, this method can also result in cumbersome expressions and, for certain protocols, more direct and efficient proof methods are available which we discuss next.

\subheading{Second Key-Rate Proof Method}
When working with a single state protocol, as mentioned, it is sufficient to consider $E$'s forward channel attack as simply biasing $B$'s measurement result.  This opens up an alternative proof strategy consisting of the following three steps \cite{SQKD-Krawec-dissertation}:
\begin{enumerate}
  \item First, consider $E$'s bias to be $0$ (see Equation \ref{eq:bias-atk}).  In this case, $E$ is actually performing the identity operator on the forward channel and, so, the protocol reduces to a one-way protocol consisting of three states.  Namely, the protocol reduces to a protocol where $B$ (who is no longer classical) prepares $\ket{\psi_0}$ or $\ket{0}$ or $\ket{1}$.  Eve attacks the reverse channel normally through a unitary probe, and $A$ performs her operations as dictated by the original protocol.  Since this is essentially a one-way protocol (as $E$ is not attacking the forward channel when the bias is $0$), its security analysis may be potentially easier (or, even, already completed in past work since, often, the protocol reduces to one that is mathematically equivalent to a known one-way QKD protocol) thus giving a bound on $S(A|E)_{\rho_0}$ in this case.
  \item Next, consider a fixed reverse attack probe but now alter the bias parameter $b$.  Let $\rho_b$ be the resulting density operator for a bias value of $b$ (where this $b$ is the actual observed bias in the operation of the protocol).  We need to compute $S(A|E)_{\rho_b}$, the conditional entropy for the actual attack we observe.  Since we know, from the first step, the value of $S(A|E)_{\rho_0}$ (i.e., when there is no bias), we may compute $S(A|E)_{\rho_b}$ using the continuity of von Neumann entropy \cite{QC-alicki2004continuity,QC-fannes-audenaert-inequality,QC-winter2016tight}.  Indeed, using a continuity bound in \cite{QC-winter2016tight}, we may write (since $\dim \mathcal{H}_A = 2$):
\begin{equation}
  S(A|E)_{\rho_b} \ge S(A|E)_{\rho_0} - \delta - (1+\delta)h\left(\frac{\delta}{1+\delta}\right),
\end{equation}
where:
\begin{equation}
\delta = \frac{1}{2}||\rho_b - \rho_0||.
\end{equation}
Thus, the goal of this second step, is to compute a bound on $\delta$ as a function only on observed noise parameters, including the observed bias $b$.  Of course when $b = 0$, we obtain $S(A|E)_{\rho_b} = S(A|E)_{\rho_0}$ as expected.  As the bias increases (e.g., as $E$'s forward channel attack becomes stronger), $\delta$ increases, thus causing $S(A|E)_{\rho_b}$ (the actual conditional entropy of the protocol operation) to decease, thus causing the key-rate to also decrease.
  \item Finally, the two steps are combined, however care must be taken in that, on step (1), the entropy $S(A|E)_{\rho_0}$ was bounded as a function of the observed noise - however on the one hand, the observed $X$ basis noise (when $B$ chooses $\R$) is a function now of both forwards and reverse attacks, whereas the entropy bound from step (1) assumes it is only in the reverse channel.  That is, $E$'s attack in the reverse may actually emit more $X$ basis noise by itself (when $b=0$) then the actual observed noise.  Therefore, to complete the proof, given $Q_X$, the observed $X$ basis noise and given the bias $b$, determine a bound on $\widetilde{Q}_X$, the noise produced only in the reverse channel by the unitary probe.
\end{enumerate}

Ultimately, the above method leads to simpler security proofs.  Step (1) is often achieved by recognizing that the protocol, when the bias is set to zero, reduces to a well known protocol such as the Three State BB84 \cite{QKD-BB84-three-state,QKD-BB84-three-state-v2} (as is the case of the Zou et al., protocol \cite{SQKD-less-than-4}) or the Extended B92 protocol \cite{QKD-B92-extended} (as is the case of the Reflection-Based protocol introduced in \cite{krawec2014restricted}).  However, the use of a continuity bound gives a worst-case result.  Indeed, the first method has, so far, always led to more optimistic results (all results have been lower-bounds, so there is no contradiction).  The first method also allows for finer-grained control of the result.  Indeed, as shown in \cite{krawec-reflect-new}, the bias can positively and negatively affect $E$'s uncertainty (as expected) - however this observation is not possible when using the second method as any bias automatically leads to a decrease in uncertainty (as $\delta$ increases); i.e., it leads to a worst-case bound.

\subheading{Third Key-Rate Proof Method}
Finally, the third method of proof involves reducing the SQKD protocol to a one-way, fully quantum protocol and then analyzing that protocol directly.  It was proven in \cite{krawec-entropic} that the BKM07 protocol can be reduced to an equivalent one-way protocol (where, now, both parties are actually fully quantum) of the following form:
\protocol{Equivalent One-Way Protocol for BKM07 \cite{krawec-entropic}}
\begin{enumerate}
  \item $B$, who is now a fully quantum user, prepares either the state $\frac{1}{\sqrt{2}}(\ket{00}_{A_1A_2}+\ket{11}_{A_1A_2})\otimes\ket{0}_B$ or the state $\frac{1}{\sqrt{2}}(\ket{000}_{A_1A_2B} + \ket{111}_{A_1A_2B})$, choosing randomly each iteration (with the same probability that he normally would have chosen $\R$ or $\MR$ respectively in the original SQKD protocol).  He then sends the $A_1A_2$ qubits to Alice ($E$ is allowed to attack both qubits simultaneously).
  \item $A$ measures both the $A_1$ and $A_2$ qubits in either the $Z$ basis or the $X$ basis, choosing randomly.
  \item $A$ discloses her choice of basis and $B$ his choice of state preparation.  If $B$ choose to prepare the GHZ state $\frac{1}{\sqrt{2}}(\ket{000}_{A_1A_2B} + \ket{111}_{A_1A_2B})$, and if $A$ chose to measure in the $Z$ basis, this iteration may be used for raw key distillation (they should share a correlated bit).  Otherwise, the iteration may be used for error estimation.
\end{enumerate}

The proof that security of this one-way fully quantum protocol implies security of the original BKM07 protocol can be found in \cite{krawec-entropic}.  A similar reduction was recently proven for a higher dimensional SQKD protocol in \cite{sqkd-high-dim}.  It is currently an open problem as to which families of SQKD protocols have a similar reduction.

\openproblem Do all SQKD protocols have an equivalent one-way protocol that they may be reduced to?

Regardless, once reduced, the one-way protocol may be analyzed through standard techniques, for instance using entropic uncertainty relations \cite{uncertainty-survey1,uncertainty-survey-2,uncertainty-survey-3}.  This then may be translated to a key-rate bound for the semi-quantum protocol.  As with the second method, this leads to a clear and concise security bound, but it does not give as optimistic a result as the first method (due, perhaps, in part to the fact that the one-way protocol affords $E$ more attack opportunities than in the actual two-way SQKD protocol, thus causing a less than optimistic bound on security to the adversary's advantage).  Indeed, while the first method described earlier can show that BKM07 can suffer $11\%$ noise tolerance \cite{QKD-Tom-Krawec-Arbitrary}, this third method shows only $6.14\%$ \cite{krawec-entropic}.  However, that is not an entirely fair comparison: the first method relied on the collection of numerous mismatched measurements (thus allowing for a tighter bound on the entropy) whereas the third method did not use any mismatched measurements - only the error rate.  It is unclear if mismatched measurements are necessary to attain this high noise tolerance and, perhaps, the $6.14\%$ as determined by this third method is actually tight for this protocol without these statistics.

\subheading{Mismatched Measurements}

As discussed, three methods of computing the key-rate of an SQKD protocol have so far been developed.  The first method, direct computation of $S(A|E)$, combined with \emph{mismatched measurements} have so far given the most optimistic results.  Mismatched measurements are a technique originally introduced in 1993 by Barnett et al., in \cite{QKD-Tom-First} for fully-quantum protocols.  Later the technique became more refined in \cite{QKD-Tom-KeyRateIncrease,QKD-Tom-KeyRateMismatchedDistill,QKD-Tom-BB84NarrowAngle} showing that substantial improvements in noise tolerance and asymptotic efficiency are possible for fully-quantum protocols with restricted resources such as the Three State BB84 \cite{QKD-BB84-three-state,QKD-BB84-three-state-v2} or the Extended B92 \cite{QKD-B92-extended} protocols - indeed for the Three State BB84 protocol, despite $A$'s inability to send the $\ket{-}$ state, noise tolerance can be as high as the standard four-state BB84 as shown in \cite{QKD-Tom-threestate1,QKD-Tom-threestate-Krawec}.

This technique of using mismatched measurements was extended in \cite{QKD-Tom-Krawec-Arbitrary} to two-way quantum channels and semi-quantum users using two bases and extended in \cite{sqkd-high-noise} for three bases.  Using this method, one may show that the noise tolerance of the BKM07 protocol, assuming $E$'s attack is symmetric (\emph{an enforceable assumption}), is as high as BB84 as discussed in the previous section.  To compute this key-rate requires looking at $18$ different measurement statistics as shown in Table \ref{tbl:mm-stats}.  Without these statistics, the current best result is based on reducing to a one-way protocol and using an entropic uncertainty bound - using such a method does not require collecting all of these statistics (it only requires looking at error rates) but the (lower-bound bound on) noise tolerance drops to $6.14\%$ \cite{krawec-entropic}.  Whether mismatched measurements are necessary for the BKM07 protocol to attain this high noise tolerance is still an open question.  Indeed, the $6.14\%$ tolerance from \cite{krawec-entropic} is only a lower bound.

\begin{table}
\centering
\begin{tabular}{|c|c|l|}
\hline
&&\\
\textbf{Error Statistics} &$\prf_{i,1-i}$ & Forward channel $Z$ basis noise\\
&$\prr_{i,j,1-j}$& Reverse channel $Z$ basis noise\\
&$\prr_{i,R,1-i}$& $Z$ basis noise when $B$ chooses $\R$\\
&$\prr_{\pm,R,\mp}$& $X$ basis noise when $B$ chooses $\R$\\
&&\\
\hline
&&\\
\textbf{Mismatched}& $\prf_{+,i}$ & Forward channel $X\rightarrow Z$ statistic.\\
\textbf{Statistics}& $\prr_{i,j,+}$ & Reverse channel $Z\rightarrow X$ statistic.\\
&$\prr_{+,R,0}$ & Loop channel $X\rightarrow Z$ statistic\\
&$\prr_{i,R,+}$ & Loop channel $Z\rightarrow X$ statistic\\
&&\\
\hline
\end{tabular}
\caption{Showing all observable statistics used in the key-rate computation for the BKM07 protocol in \cite{QKD-Tom-Krawec-Arbitrary}.  Here, $i,j \in\{0,1\}$ and we use $\prf_{i,j}$ to denote the probability that $B$ observes $\ket{j}$ given that $A$ initially sent $\ket{i}$ and $B$ chooses $\MR$; $\prr_{i,j,k}$ is the probability that $A$ observes $\ket{k}$ (for $k\in\{0,1,+,-\}$) conditioned on $A$ initially sending $\ket{i}$, $B$ choosing $\MR$ and actually observing $\ket{j}$, and finally $A$ choosing to measure in the correct basis to observe $\ket{k}$; finally $\prr_{i,R,k}$ is similar, but now conditioning on $B$ choosing $\R$.  In \cite{sqkd-high-noise}, this was extended to allow the quantum user to choose from three bases, $Z$, $X$, or $Y$; while this increases noise tolerance, it also roughly doubles the number of statistics needed for mismatched measurements.  Note that by ``Loop channel'' above, we mean the joint channel when $B$ chooses $\R$.}\label{tbl:mm-stats}
\end{table}

\openproblem Are mismatched statistics necessary for the BKM07 protocol to attain the same noise tolerance as BB84, namely $11\%$?  Or can this tolerance be achieved by looking only at error statistics?

Note that we did not ask the question in regards to \emph{any} semi-quantum protocol.  In fact, it was shown in \cite{sqkd-limited-measure,sqkd-classical-quantum} that for some SQKD protocols (specifically the two developed in those references), mismatched measurements are necessary to attain any level of security.  That is, \emph{without mismatched measurements, there are semi-quantum protocols that are completely insecure.}  This seems to suggest that mismatched measurements may be necessary for all SQKD protocols (either to show any form of security or, in the case of BKM07, to improve security bounds) though an exact proof of this is elusive.  It is interesting to note, however, that by dropping the resource requirements of users (namely, when moving from the fully quantum to the semi-quantum setting), one can use additional classical post processing (e.g., mismatched measurements), to compensate.  This seems to suggest the semi-quantum model of communication can shed light on interesting fundamental connections between classical and quantum information processing.

Overall, as of writing this, several SQKD protocols have a key-rate analysis lower-bound thus giving us a lower-bound on the protocol's noise tolerance.  A summary of the current best case noise tolerances are shown in Table \ref{tbl:noise-tol}.  Noise tolerances are reported here based on the $Z$ basis error in the forward and reverse channel, denoted here as $Q$ (we also assume $Q_X = Q$).  In particular, the value reported is the maximal $Q$ for which the resulting key-rate $r$ is positive.  Note that many key-rate proofs for SQKD protocols support different noise scenarios, including different $Z$ basis noise rates in the forward and reverse channels; we only report the symmetric case here for simplicity in presentation.  Where appropriate we also assume depolarization channel noise.  Complete details for alternative scenarios, if available, can be found in the original reference for the proof of security provided in the table.

\begin{table}
\centering
\begin{tabular}{|r|c|c|l|}
\hline
Original Protocol & Noise & Proof & Comments\\
& Tolerance & Reference &\\
\hline
BKM07 \cite{SQKD-first} & $11\%$ & \cite{QKD-Tom-Krawec-Arbitrary}&\\
Single State by Zou et al., \cite{SQKD-less-than-4}  & $9.65\%$ & \cite{zhang2018security}&\\
Reflection-Based \cite{krawec2014restricted} & $5.36\%$ & \cite{krawec-reflect-new}&\\
Semi-Quantum B92 \cite{sqkd-single-state-b92} & $3.46\%$ & \cite{sqkd-single-state-b92}&\\
Single-$A$-Measurement \cite{sqkd-limited-measure}  & $11\%$ & \cite{sqkd-limited-measure} & MM Required\\
Classical-to-Quantum \cite{sqkd-classical-quantum} & $<1\%$ & \cite{sqkd-classical-quantum} & MM Required; Noise\\
&&&tolerance depends on\\
&&&distance from classical.\\
High-Noise-SQKD M2 \cite{sqkd-high-noise} & $16.4\%$ & \cite{sqkd-high-noise}&\\
High-Noise-SQKD M3 \cite{sqkd-high-noise}  & $26\%$ & \cite{sqkd-high-noise}&\\
High-Dimensional SQKD \cite{sqkd-high-dim} & $30\%$ & \cite{sqkd-high-dim} & Noise tolerance increases\\
&&&to $30\%$ as dimension\\
&&&approaches infinity\\
\hline
\end{tabular}
\caption{Showing state of the art best noise tolerances for those SQKD protocols which have this analysis performed.  ``MM Required'' means that mismatched measurements are required for the protocol to be secure at all (i.e., the protocol is completely insecure without them); note that mismatched measurements may be used in the above results for other protocols besides those specifically marked as such, but it is not required for security - see text for discussion.}\label{tbl:noise-tol}
\end{table}

\section{Multi-User SQKD}

While the vast effort in SQKD research (i.e., research specific to key distribution in the semi-quantum model) is in trying to discover, and prove secure, novel protocols requiring fewer resources on the part of the users, other directions have also seen great interest.  Perhaps the most fruitful as of writing this is the development of multi-user protocols.  Multi-user protocols within the semi-quantum realm come in two flavors: first is trusted quantum user (where this quantum user is trusted and, generally, shares the secret key) and the second is the mediated model (where the quantum user is adversarial and should not share the key).

The first multi-user SQKD protocols were introduced independently in \cite{cl-A} and \cite{multi1}.  The network topology assumed by their protocols is circular in that users communicate in sequence.  Here, one fully quantum user, who is trusted and is one of the key-holders, transmits quantum resources to the first classical user $B_1$.  This user then can perform some semi-quantum operation (e.g., $\MR$ or $\R$), forwarding a qubit to the next classical user $B_2$.  This repeats for the next user and so on until $B_n$ at which point the qubit returns to the quantum user who is free to measure in any basis.  After the protocol, all $B_i$'s transmit their choice of operation and $A$ will transmit her basis choice.  It is assumed this broadcast communication is done in an authenticated manner, though the classical communication mechanism required for this to operate successfully is not discussed.  In general, whenever two $B$'s choose $\MR$ those users share a key bit.  Key bits are shared with the quantum user $A$ whenever a $B_i$ chooses $\MR$ and the quantum $A$ chooses to measure in the $Z$ basis.  Thus, these protocols permit different subgroups of users to share different keys.



\begin{figure}
  \centering
  \includegraphics[width=200pt]{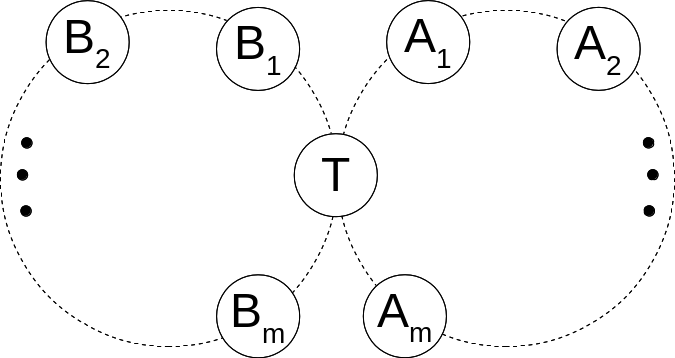}
\caption{Based on an image in \cite{multi3} showing the assumed network topology for their protocol in that reference.  Each $B_i$ and $A_i$ is a classical user while $T$ is a fully quantum user.}\label{fig:multi-user-net-alt}
\end{figure}

Other multi-user protocols in the semi quantum model have also been proposed.  In \cite{multi2} a protocol based on a trusted server preparing GHZ states was described and security analyzed with regards to certain attacks including some attacks based on an adversarial server.  One of the main limitations to previous multi-user protocols is that, for $m$ users to agree on a key, all $m$ have to choose the ``correct'' options for that event to happen (e.g., all classical users must choose $\MR$ in the protocol of \cite{multi1}).  To improve efficiency, a new multi-user protocol was proposed in \cite{multi3} which uses \emph{cluster states} \cite{cluster1} and an alternative network topology shown in Figure \ref{fig:multi-user-net-alt}.  Their protocol allowed for a roughly quadratic speedup in efficiency over previous work.  Also provided in \cite{multi3} was an information theoretic analysis of the key-rate showing a maximal noise tolerance of $2.82\%$.

\begin{figure}
  \centering
  \includegraphics[width=225pt]{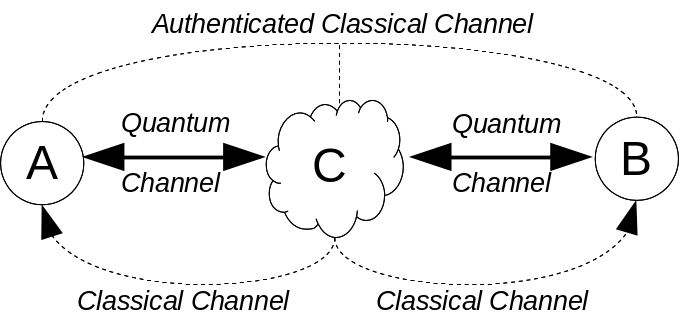}
\caption{Based on an image in \cite{sqkd-med-first} showing the structure of a mediated SQKD protocol.  Here a central, fully quantum, server $C$ (which may be adversarial) prepares and measures quantum states.  $A$ and $B$ are classical users.  An authenticated classical channel connects the two users while a standard (unauthenticated) classical channel connects the server to each user.}\label{fig:med}
\end{figure}

Beyond multi-user protocols where $M$ users all wish to agree on a key, an alternative model involving multiple users is the \emph{mediated semi-quantum} model.  This model was first introduced in 2015 in \cite{sqkd-med-first} and it involves a fully quantum server and two ``classical'' users $A$ and $B$.  These two users wish to agree on a secret key known only to them and \emph{not} the server.  Furthermore, they do not trust the server who may even be adversarial.  Two forms of adversarial models were considered: first a semi-honest server who follows the protocol but may attempt to learn additional information later; alternatively a stronger fully adversarial model is also considered.  A general scenario of this framework is shown in Figure \ref{fig:med}.  The original mediated protocol consisted of the following steps:

\protocol{Mediated SQKD \cite{sqkd-med-first}}
\begin{enumerate}
  \item The server prepares a Bell state $\ket{\Phi^+} = \frac{1}{\sqrt{2}}(\ket{00}+\ket{11})$, sending one particle to $A$ and the other to $B$.
  \item Each user, $A$ and $B$, choose, independently, to either $\MR$ or to $\R$.
  \item When both particles return to the server, the fully quantum server performs a Bell measurement, sending the classical message ``$-$'' if and only if the outcome is $\ket{\Phi^-}=\frac{1}{\sqrt{2}}(\ket{00}-\ket{11})$.  For any of the other three potential outcomes, the server sends the classical message ``$+$''.
  \item $A$ and $B$ both divulge their choice of operation.  If the server sent the message ``$-$'' and if both users chose $\MR$, they will use their measurement results as their raw key.  Note that if both users chose $\R$, the server should always send the message ``$+$'' and any other result is counted as an error.
\end{enumerate}

Proof of security in \cite{sqkd-med-first} (improved in \cite{sqkd-med-improved}) assumes an adversarial server may prepare any arbitrary state on step (1) (possibly entangled with its quantum ancilla) and, furthermore, may perform any quantum operation on step (3).  Furthermore, there may be third-party adversaries attacking the quantum channel and the classical communication between the server and the users (the classical channel connecting $A$ and $B$ needs to be authenticated, however the classical channel between the users and the server is not authenticated and so subject to manipulation by an adversary - security is still possible).  This shows that even with classical capabilities, users may enforce security of a more powerful quantum server.  Furthermore, as shown in \cite{sqkd-med-improved}, the noise tolerance can approach $22.05\%$ if the server is semi-honest or $13.04\%$ if the server is fully adversarial.  As shown in \cite{mm-sqkd}, if two independent mediators are used by $A$ and $B$ (referred to in that source as the \emph{multi-mediated SQKD model}), noise tolerance can increase to $18.7\%$ if both servers are adversarial but do not collude with each other (compared to $13.04\%$ with only one server).

Several other mediated protocols have since been introduced, mostly with the goal of providing greater efficiency (indeed, the original mediated protocol is very inefficient with many iterations being lost due to incompatible choices or measurement results), or fewer resource requirements on the users or server.  In \cite{sqkd-med-no-measure}, a mediated SQKD protocol was proposed where classical users did not have to measure (assuming perfect qubit channels) but, instead, could choose to $\R$ or $\Perm$.  This protocol also had greater efficiency than the original mediated SQKD protocol.  The authors of \cite{med-single-photon} developed a new mediated SQKD protocol where the server needs to only send single photons and perform single photon measurements.  Users must choose either $\R$ or $\MR$.  This protocol decreases the quantum complexity of the server, a useful direction to move towards as this mediated model may prove to be a practically beneficial quantum communication infrastructure.  Other ``light-weight'' mediated protocols were presented in \cite{med-light,med-light-2} designed to help mitigate trojan horse attacks against the classical users.  Finally, in \cite{med-prac}, a new mediated protocol was designed where users need only to \texttt{Measure} (\emph{but not resend}; thus users do not need a single photon source for this protocol) or $\R$.  This protocol, which was also experimentally implemented, shows that the mediated model of SQKD is a practical possibility.  We will discuss this protocol in more detail when we turn our attention to practical issues of semi-quantum cryptography.

\section{Beyond Key Distribution}

The original application of semi-quantum cryptography, much like standard, fully-quantum cryptography, was to solve the key distribution problem.  Investigating this makes sense as it is a much celebrated result showing a very clear advantage to quantum communication over classical communication (key distribution using only classical communication, as mentioned in the introduction, requires one to make computational assumptions on the adversary's capabilities).  However, the semi-quantum model of communication, involving at least one fully quantum party and one (or more) limited, ``classical'' parties, can be, and has been, applied to other problems.

\subsection{Secret Sharing}

Perhaps the first application of semi-quantum communication outside of the key-distribution problem was to the task of \emph{secret sharing} \cite{secret-orig,secret-survey}.  Secret sharing is a primitive used in numerous other cryptographic protocols and consists of a \emph{dealer} (who has some secret $s$) and $n$ other parties.  The dealer creates $n$ shares of this secret and sends one share to each party (there are $n$ parties).  In its simplest form, it should be that if $t$ or more parties come together with their respective shares, the original secret may be recovered; however if one has strictly less than $t$ shares, the secret cannot be learned.  While information theoretic secret sharing is possible using classical communication only, one of the advantages to using quantum protocols is the additional ability to detect eavesdropping \cite{secret-quantum-first,secret-quantum-second} or to potentially decrease share size (thus increasing communication efficiency) \cite{secret-quantum-third}.  Alternatively, quantum protocols must be used if the original secret itself is quantum.

The first semi-quantum secret sharing (SQSS) protocol was developed in 2010 by Qin Li et al., in \cite{sqkd-secret-first}.  Here the dealer, $A$, is quantum while two parties $B$ and $C$ are both classical.  The threshold $t$ is set to $2$ (thus both $B$ and $C$ must come together to recover the secret) and the secret itself is classical data.  Two protocols were presented, one requiring the $\Perm$ operation, the other using only $\R$ and $\MR$.  We present the second here as it is easier to follow:

\protocol{First Semi-Quantum Secret Sharing Protocol \cite{sqkd-secret-first}}
\begin{enumerate}
  \item $A$, the fully-quantum dealer who holds the secret $s$ (a bit string), creates $N$ GHZ states, each of the form:
\begin{equation}\label{eq:secret-state}
\ket{\psi_0} = \frac{1}{\sqrt{2}}(\ket{+++} + \ket{---}).
\end{equation}
She sends one particle to $B$, another to $C$, and keeps the third private in her own lab.

\item When each user receives a qubit, parties choose either to $\MR$ or to $\R$.

\item When the qubits return to $A$, she stores them and alerts $B$ and $C$.  The two parties then disclose their choice of operation for each qubit.

\item For each of the $N$ triplets, and based on $B$ and $C$'s choice, $A$ performs the following operations:
\begin{itemize}
  \item Case 1: $B = \MR$ and $C = \MR$.  Then $A$ measures her qubit in the $Z$ basis.
  \item Case 2: $B = \MR$ and $C = \R$.  Then $A$ performs a Bell measurement with her qubit and $C$'s reflected qubit.
  \item Case 3: $B = \R$ and $C = \MR$.  Then $A$ performs a Bell measurement on her qubit and $B$'s qubit.
  \item Case 4: $B = \R$ and $C = \R$.  Then $A$ performs an appropriate three-qubit measurement where one basis state is $\frac{1}{\sqrt{2}}(\ket{+++}+\ket{---})$.
\end{itemize}
Cases 2, 3, and 4 are used only for error detection (along with a random subset of Case 1 instances).  Case 1 produces a bit string $k_A$ for the dealer $A$ of size $n < N$ (the length of $n$ is expected to be $N/4-m$ bits where $m$ is the size of the random subset used for error detection in the Case 1 instances).  The dealer then sends $s \oplus k_A$ to $B$ and $C$.
\end{enumerate}

The claim for correctness and security is that the string $k_A$ is random and independent of $B$ and $C$'s individual information.  However, $B$ and $C$ can only recover $k_A$ by XOR'ing their measurement results.  This can be seen by rewriting Equation \ref{eq:secret-state} in the $Z$ basis:
\[
\ket{\psi_0} = \frac{1}{\sqrt{2}}\left(\ket{0}_A\otimes\frac{\ket{00}_{BC} + \ket{11}_{BC}}{\sqrt{2}} + \ket{1}_A\otimes\frac{\ket{01}_{BC}+\ket{10}_{BC}}{\sqrt{2}}\right).
\]
Note that $B$ and $C$'s bits are randomly distributed and that by XOR'ing their results, they recover $A$'s bit.

In \cite{sqkd-secret-second}, a SQSS protocol was proposed which only required the dealer to prepare $N$ copies of the state $\frac{1}{\sqrt{2}}(\ket{+,0} + \ket{-,1})$, sending one particle to $B$ and the other to $C$.  Both these first papers \cite{sqkd-secret-first,sqkd-secret-second} showed secret sharing is possible with semi-quantum users, however it required the generation of entangled states (e.g., Equation \ref{eq:secret-state}) and the protocol only worked with two parties.  In \cite{sqkd-secret-multiparty}, both issues were considered and improved on.  First, a protocol was proposed for $n$ classical parties, extending the technique of \cite{sqkd-secret-first}.  For this to operate, the dealer must prepare a state of the form:
\[
\frac{1}{2^{(n+1)/2}}\left(\ket{+}^{\otimes n+1} + \ket{-}^{\otimes n+1}\right).
\]
Exact details of the protocol may be found in \cite{sqkd-secret-multiparty}, however, this still requires the generation of a highly entangled state.  To mitigate this, the same authors in \cite{sqkd-secret-multiparty} propose a two-party SQSS protocol where the dealer need only prepare separable states of the form $\ket{+}\ket{+}$, sending one particle to $B$ and the other to $C$.  For the multiparty case, the dealer must prepare the $n$ qubit state $\ket{+}^{\otimes n}$ thus creating a more practical system (though, as pointed out in \cite{sqkd-secret-multiparty}, efficiency of the protocol may be problematic for large $n$).

So far these SQSS protocols shared a secret through the use of a randomly generated pad.  That is, before the protocol executed, the dealer had no way to deliberately create shares based on the secret itself.  An alternative SQSS protocol was devised in \cite{sqkd-secret-message} where the actual creation of shares depends on the secret - thus, there was no need for an additional transmission of $s \oplus k_A$ as was required with these other protocols discussed so far.  In their protocol, the secret is a single bit $b \in \{0,1\}$ (though this may be increased simply by running multiple instances of the protocol in sequence).  At the start, $A$ prepares $N$ states of the form $\frac{1}{\sqrt{2}}(\ket{+++} + (-1)^b\ket{---})$ (thus $N$ quantum states are required for a secret of one bit).  One particle is sent to $B$, another to $C$, and a third is kept private.  At the end, the secret bit can only be recovered if all three users (including the dealer in this protocol) present their final classical shares, distributed during the quantum stage of the protocol.  In the same paper, an $n$ party protocol was also developed, though still requiring $N$ quantum states per classical bit of the secret and requiring the generation of entangled states.  In \cite{sqkd-secret-message-atk} an intercept-resend attack was shown against this protocol allowing a participant to recover the message without having to collaborate.  While a fix was presented in that paper, it required parties to be fully quantum (in that they should also measure in the $X$ basis).  However, an alternative fix was presented in \cite{sqkd-secret-message-atk-2} which is semi-quantum.

Numerous other SQSS protocols have been proposed in addition to these.  In \cite{sqkd-secret-limited}, a new SQSS protocol was proposed using higher-dimensional states that affords greater efficiency.  A ``circular'' SQSS protocol was developed in \cite{sqkd-secret-circle} which only required single particles and removed the need for measurements (though it does require the $\Perm$ operation).  This protocol is ``circular'' in its network topology, requiring these single particles to travel from the dealer $A$ to $B$, then to $C$, and finally return to the fully-quantum $A$.

An SQSS protocol without the need for measurement was proposed in \cite{sqkd-secret-limited-2}.  A $d$-dimensional protocol was proposed in \cite{high-dim-secret-share} which also did not require classical users to measure and supported multiple (beyond two) classical users.  A secret sharing protocol using $W$-states for encoding (as opposed to Bell states or GHZ states) was developed in \cite{secret-W-state}.

In \cite{sqkd-secret-convert} a new multi-user (i.e., where the number of parties was greater than two) protocol was developed with greater efficiency than prior multi-user versions at the time of its publication; also it was proven in that reference that multi-user SQSS protocols may be converted to SQKD protocols and a construction was given (furthermore, some simulations were performed on the IBM quantum computer).  Bell states were used to create an SQSS protocol in \cite{sqkd-secret-bell} (where classical users applied $\R$ or $\MR$) though a security flaw was found in \cite{sqkd-secret-bell-atk} (no fix was provided leaving this an open question); an alternative SQSS protocol using Bell states was developed in \cite{sqkd-secret-bell-2} though where classical users need also the $\Perm$ operation.  An interesting encoding scheme for SQSS was developed in \cite{sqkd-secret-arb} allowing a secret to be shared by $A$ preparing multiple, initially unspecified, entangled states (their protocol also works for more than two classical users); though in \cite{sqkd-secret-arb-atk} an attack was found on this protocol, however possible fixes were also presented.  A scalable SQSS protocol was developed in \cite{sqss-scale} allowing users to be added or removed by the dealer.



One interesting observation is that all current SQSS protocols have the dealer as the fully-quantum user.  This makes sense from a practical standpoint (it should be the dealer who has the most capabilities).  However, from a theoretical stand-point can one construct other scenarios?

\openproblem Does there exist an SQSS protocol where the dealer is classical?  There are two possible variants: first, one of the participants is fully quantum and ``helps'' by getting the protocol started (e.g., sending quantum resources to the classical dealer).  A second is in line with mediated SQKD protocols as discussed earlier: namely, the dealer and all participants are classical, but there is an untrusted quantum server to perform the needed quantum operations.  Showing protocols exist for both settings would be an interesting theoretical result; showing a protocol in the mediated case may also be interesting from a practical standpoint as one could envision a future communication infrastructure where untrusted servers help facilitate both key distribution (through mediated SQKD protocols) and other cryptographic protocols (such as secret sharing) performed by classical users.

Finally, does sharing a \emph{quantum} state make sense in the semi-quantum setting?  In \cite{sqss-quantum}, the authors proposed a protocol where a quantum state may be shared between classical $A$ and quantum $B$ (the dealer is also quantum and, of course, only the quantum user can recover the secret later).  Further research in this may prove interesting.

\subsection{Secure Direct Communication}

Secure direct communication (SDC) is the task of sending a message directly from $A$ to $B$, through a quantum channel, without having to first establish a shared secret key (beyond that needed for authentication of classical information).  SDC protocol development in the fully-quantum model dates back to the early 2000's and there have been several protocols since with various advantages and disadvantages (see \cite{fully-sdc0,fully-sdc00,fully-sdc1,fully-sdc2,fully-sdc3} for just a few instances in the fully-quantum setting).

The first semi-quantum SDC protocol was developed in \cite{sqdc-first} showing that, like with key distribution, the task of SDC is also possible in the semi-quantum model.  In their protocol, the sender of the message $m \in \{0,1\}^n$ is the classical user ($B$) while the receiver is the quantum user ($A$).  Their protocol utilizes a hash function $h:\{0,1\}^n \rightarrow\{0,1\}^k$ for some $k < n$.  The protocol operates as follows:

\protocol{First Semi-Quantum SDC Protocol \cite{sqdc-first}}
\begin{enumerate}
  \item $A$ prepares $N \approx 4(n+k)$ qubits, each of which is prepared independently at randomly as one of the four states $\ket{0}$, $\ket{1}$, $\ket{+}$, or $\ket{-}$.  She sends these qubits to the classical user $B$ (who is the message sender).
  \item When $B$ receives the qubits, half are selected randomly for error testing, the other half for message encoding.  Those selected for error testing are subjected to a random choice of $\R$ or $\MR$.  When these qubits return to $A$, $B$ informs her of their indices (he is still, through a delay line, holding on to the other half of the qubits) allowing parties to check for eavesdroppers in the standard way.  If this is detected, parties immediately abort.
  \item Assuming no eavesdropping was detected on the test half, $B$ will choose $n+k$ random qubits from the remaining portion and measure them in the $Z$ basis.  He then computes the classical bit string $\hat{M} = M || h(M)$ where $||$ represents bit-string concatenation.  Finally, for each measured qubit and for each bit in $\hat{M}$ he will prepare a new $Z$ basis qubit either in the same state he measured if that bit of $\hat{M}$ is $0$ or he will prepare the opposite $Z$ basis state if the bit of $\hat{M}$ is $1$.  All qubits (both those he measured and encoded the message and hash in and those others he is choosing $\R$), are returned to the quantum user (the receiver of the message).  Notice that, due to the random choice of qubit preparation, the message, at this point, is encoded using a classical one time pad which only $A$ and $B$ know the key to.
  \item $A$ receives all qubits and is told from $B$ which are those he encoded his message in and which were reflected.  Security is verified on all reflected qubits; for the others, $A$ measures in the $Z$ basis to receive the classical string $\hat{M}' = M'||H'$ and verifies that $H' = h(M')$ ensuring that an adversary did not tamper with the message.
\end{enumerate}

Note that, in its original form as described above, the protocol was actually shown in \cite{sqdc-attack} to be susceptible to a Double CNOT attack; two potential solutions were presented in that reference, however, such as changing the protocol so that $B$ uses the $\Perm$ operation before sending any qubits back to the quantum user.

Other semi-quantum SDC protocols have been proposed.  While the above protocol from \cite{sqdc-first} allows a classical user to send a message to a quantum user, the reverse direction, namely sending a message from the quantum user to the classical user, was considered in \cite{sqdc-2} where a novel protocol was developed allowing for this functionality.  A protocol utilizing EPR pairs allowing a classical user to send a message to the quantum user was developed in \cite{sqdc-bell,sqdc-bell-2}.  Two protocols were proposed in \cite{sqdc-two-eff} which also used Bell states though claimed higher qubit efficiency.

Another protocol developed in \cite{sqdc-auth} removed the need for an authenticated channel by assuming a pre-shared secret key is first agreed on (though, this key must be linear in the size of the message).  A so-called \emph{delay attack} on these authenticated style protocols was discovered in \cite{sqdc-auth-atk} along with a new protocol to counter it (this new protocol also had the added advantage that it required less resource requirements on the part of the classical user).  This was further improved in \cite{sqdc-auth-eco,sqdc-auth-eco-2} which reduced the required resources on the part of the quantum user also, though added the requirement again of an authenticated classical channel.  An authenticated SDC protocol using only single qubits was proposed in \cite{sqkd-sdc-auth-single}.  Finally, \cite{sqdc-auth-two} proposed two new SDC protocols allowing quantum $A$ to send a message to classical $B$ in such a way that both users can verify the authenticity of the message (assuming a pre-shared key was already shared) using quantum error correction codes.

The notion of \emph{Quantum Dialog}, first introduced for fully quantum users in \cite{qkd-qd-0,qkd-qd-1}, is similar to SDC except that it allows for a message to be transmitted from $A$ to $B$ and a separate message from $B$ to $A$.  This was extended to the semi-quantum domain first in \cite{sqdc-qd-1} (which also proposed a novel semi-quantum SDC protocol) and an alternative protocol in \cite{sqdc-qd-2}.  A quantum dialog protocol consisting of two classical users and an untrusted server was presented in \cite{sqdc-qd-auth}; their protocol could also tolerate certain noisy channels.

\subsection{Other Cryptographic Protocols}

While secret sharing and secure direct communication seem to be the two largest avenues of research in semi-quantum cryptography, outside of key distribution, other cryptographic primitives have recently begun to be investigated.

One avenue, similar to key distribution, is quantum key agreement \cite{QKA-1,QKA-2,QKA-3}.  Here, the goal is to ensure that both $A$ and $B$ contribute to the generated raw key equally and that no one party can bias the result.  Protocols achieving this in the semi-quantum case have been proposed in \cite{sqdc-qd-1,sqkd-qka-0,sqkd-qka-1,sqkd-qka-2}.

Private state comparison is a cryptographic primitive where parties $A$ and $B$ each hold some data $i_A$ and $i_B$ respectively (e.g., parties hold two numbers) and they wish to compare their data to determine, for instance, who has the larger number or, in the case of private state comparison, whether they are equal or unequal.  However, they wish to do so in a way that does not reveal their data to the other party.  This is a particular instance of Secure Multiparty Computation (SMC), an important area of research in cryptography \cite{smc1,smc2}.   This task has been extended to the quantum domain through several works \cite{qpc1,qpc2,qpc3} (this is hardly an exhaustive list of fully-quantum results - see \cite{qpc-survey} for a review); of course, Lo \cite{qpc-imp} proved that the equality function cannot be computed securely even using quantum means.  Thus, research in this area often involves the use of a third party or weaker security models.  Recently, and relevant to us, this task has been extended to semi-quantum communication.

The first semi-quantum private comparison (SQPC) protocols were developed independently in \cite{sqpc-first,sqpc-first-2} where the two users $A$ and $B$ were classical but the third-party was fully quantum.  $A$ holds classical data $M_A$ and $B$ holds $M_B$; parties wish to know if $M_A = M_B$ without $A$ learning $M_B$ or $B$ learning $M_A$ (also, the third party should not learn either $M_A$ or $M_B$).  To give some idea how these protocols operate, we present the main details from the protocol in \cite{sqpc-first}:

\protocol{SQPC Protocol \cite{sqpc-first}}
\begin{enumerate}
  \item Parties $A$ and $B$, using the fully quantum third-party, first run a mediated SQKD protocol (such as the one in \cite{sqkd-med-first}) to establish a shared secret key which only $A$ and $B$ know, but not the third party.  Call this key $k_{AB}$.
  \item Next, each party separately establishes a private key with the third party using a standard SQKD protocol (e.g., BKM07).  Call these keys $k_{AT}$ (held by $A$ and the third party) and $k_{BT}$ (between $B$ and the third party).
  \item The quantum third party prepares a sufficient number of Bell states, choosing randomly from all four possibilities.  One particle of each pair is sent to $A$ and the other to $B$.  These parties then, independently, choose either to $\MR$ or to $\R$.
  \item For each returning Bell state pair, the third party performs a Bell measurement on them.  If the result was the same Bell state that was initially prepared, the third party sends the classical message ``0'' to both parties; if the Bell state observed is different (due to one party choosing $\MR$ for instance), then the third party sends the message ``$1$.''
  \item $A$ and $B$ disclose their choices and run a suitable error-checking protocol comparing their measurement results and the third party's response on a suitably chosen random subset of states.  On all other iterations where both parties choose $\MR$, they now share a correlated string $K_A$ and $K_B$ (the third party discloses his initial Bell state preparation allowing $B$ to ``flip'' the correct bits of $K_B$ so that $K_A = K_B$).
  \item $A$ sends the message:
\[
C_A = M_A \oplus K_A \oplus K_{AB} \oplus K_{AT},
\]
while $B$ sends:
\[
C_B = M_B \oplus K_B \oplus K_{AB} \oplus K_{BT}
\]
to the third party
  \item Finally, the third party computes $C_A \oplus C_B \oplus K_{AT} \oplus K_{BT}$ and announces the result.  Note that if this is the zero string then $M_A = M_B$; otherwise $M_A \ne M_B$.
\end{enumerate}
A security analysis and also the effects of noise, was performed on this protocol in \cite{sqpc-first}.

Other SQPC protocols have been proposed.  In \cite{sqpc-single,sqpc-single-2}, protocols requiring only single photons were presented.  Another protocol in \cite{sqkd-qka-2} was developed which used a new semi-quantum key agreement protocol developed in the same reference.  A protocol where the quantum third party was not required to prepare entangled states was developed in \cite{sqpc-no-ent}.

Semi-quantum protocols for identity verification were developed recently in \cite{sqkd-ident,sqkd-ident-2}.  These protocols allow quantum $A$ and classical $B$ to verify their identities assuming a pre-shared secret key.  In \cite{database}, a protocol was developed allowing a classical user to securely query a database entry owned by another classical user.  Here, the database owner should not know the query and the user asking should not learn anything else about the database.  This protocol required a quantum third party of course.

Finally, a form of measurement device independent protocol was constructed in \cite{sqkd-mdi}.  Here a quantum $A$ sends qubits to both the third party measurement device and to classical $B$.  In this protocol $B$ is allowed to $\R$ (in this case reflecting to the third party measurement device) or discard the qubit, preparing a fresh $Z$ basis state (since $B$ does not measure, he cannot perform $\MR$ exactly thus he is dropping the qubit from $A$ and preparing a fresh one independent of the state received).  The third party measurement device must perform a Bell measurement.  Also, an oblivious transfer (OT) protocol was presented in \cite{sqkd-ot} and a quantum signature scheme developed in \cite{sqkd-sig}.  More research in device independence, along with alternative cryptographic primitives (such as OT or signatures, perhaps using alternative security models such as bounded storage \cite{bounded-storage,bounded-storage2} or noisy storage \cite{noisy1,noisy2}) for semi-quantum protocols would be highly valuable.

\section{Practical Semi-Quantum}

While the original motivating factor behind the semi-quantum model of communication is to study the theoretical question ``how quantum must a protocol be to gain an advantage over its classical counterpart'' \cite{SQKD-first}, as QKD technology matures, it is worth also considering the question: can practical SQKD systems be implemented?  Indeed, in the fully-quantum setting (e.g., BB84), companies already exist producing commercial QKD systems and QKD has been used in several real-world applications.  Outside of these applications, there continues to be rapid progress in experimental research involving QKD systems.  For a general survey of fully-quantum cryptography, the reader is again referred to \cite{qkd-survey4}.

When it comes to implementing a semi-quantum protocol, several major challenges quickly arise.  First, semi-quantum protocols require a two-way quantum channel.  Second, many theoretical protocols require the classical user to $\MR$ - in practice this would be implemented through a photon detector which absorbs the photon and, so, to ``resend'' $B$ would need to prepare a fresh photon opening the door to multiple attacks \cite{tag-attack,tag-atk-counter,tag-atk-2}.  Third, the act of switching between $\MR$ and $\R$ requires fast, low noise, switching capabilities.  Finally, device imperfections need to be considered and finite-key security proofs must be derived.

As it turns out, the first major challenge, the dependence on a two-way quantum channel, may not be as much a hindrance as initially one might think and may, in fact, be advantageous in some scenarios.  Indeed, several fully-quantum QKD systems, especially in the continuous variable (CV) model \cite{CV1,CV2,CV3,CV4,CV5,CV6,CV-floodlight}, have been proposed and experimentally implemented, using a two-way channel and, furthermore, have shown in some cases to hold an advantage to one-way quantum communication in terms of noise tolerance \cite{CV1} or efficiency \cite{CV-floodlight}; they are also potentially more secure against source preparation noise \cite{CV3}.  It would be interesting to see if these CV techniques could be applied to the semi-quantum scenario.  Of course, for this, the notion of ``semi-quantum'' must be defined for continuous variables.

\openproblem Can a rigorous definition of continuous variable semi-quantum communication be developed? What kinds of protocols can be discovered in such a setting and what are their advantages, especially with regards to two-way quantum communication?

For the semi-quantum case, it has been shown, at least in the ideal theoretical perfect qubit case, that two-way channels can be used advantageously to promote the noise tolerance of protocols \cite{sqkd-high-noise}.  While this is the perfect qubit scenario, it does show that two-way channels can be advantageous for semi-quantum communication.  Furthermore, the techniques there may perhaps be applied to practical SQKD systems.

The second major challenge is perhaps the most critical to overcome.  If $B$ prepares fresh qubits after performing a measurement, this opens the system to photon tagging attacks \cite{tag-attack} or trojan horse attacks \cite{tag-atk-2}.  Thus, for any SQKD protocol to be practical, it would seem that $B$ should never prepare a fresh photon when performing the theoretical $\MR$ operation.  As it turns out three SQKD protocols \cite{sqkd-mirror,sqkd-reflect-prac,med-prac}, so far, have been proposed which are able to choose $\R$ and $\MR$ yet, when choosing the latter, do not actually result in a new photon being created.

The first protocol to achieve this is the so-called \emph{mirror protocol} and it was the first SQKD protocol designed with practical implementation issues in mind \cite{sqkd-mirror}.  To describe the protocol, we require the use of the Fock basis, where, briefly, we write $\ket{i,j}$ to mean a state consisting of $i$ photons in the $\ket{0}$ state and $j$ photons in the $\ket{1}$ state (physically, these may be polarization, time-bin, spatial encoding, or some other encoding as needed by the protocol).  We write $\ket{i,j}_X$ to mean a similar thing but now in the $X$ basis. $B$'s allowed operations were then refined to allow the classical user to only measure $\ket{0}$ states, ignoring $\ket{1}$ states; only measure $\ket{1}$ states, ignoring $\ket{0}$ states; or measure both $\ket{0}$ and $\ket{1}$ states.  However he does not need to ``prepare'' or ``resend'' a photon which is critical for practical SQKD security.  This operation can be done in a classical manner through the use of time-bin encoding.  For instance, to observe only photons in the $\ket{0}$ state, $B$ needs to be able to detect photons in time bin $t_0$ while reflecting the photons in time bin $t_1$ (the $\ket{1}$) states.  This requires the use of a controllable mirror (hence the name ``mirror protocol'').  The protocol, which is a single-state protocol, operates as follows:

\protocol{Mirror Protocol \cite{sqkd-mirror}}
\begin{enumerate}
  \item Fully-quantum $A$ sends a single photon in the $\ket{+}$ state which, in Fock notation, is $\ket{1,0}_X = \frac{1}{\sqrt{2}}(\ket{0,1}+\ket{1,0})$.
  \item $B$ chooses randomly either to $\R$ or to \texttt{Measure}.  If he chooses the latter, he chooses one of three options: \texttt{Measure-All}, \texttt{Measure-0}, or \texttt{Measure-1}.  These operations are described in the text above.  If he chooses one of these measure operations, he records whether he received a ``click'' (a detection) or not.  Note that if he chose, say, \texttt{Measure-0}, and assuming the state arriving at his lab is the correct $\ket{1,0}_X = \frac{1}{\sqrt{2}}(\ket{1,0}+\ket{0,1})$, he will only see a click with probability $1/2$.  If he does not see a click, the state is projected to the unobserved state (e.g., the other time bin).
  \item $A$ measures the returning state in either the $Z$ or $X$ basis, choosing randomly.
  \item Following her measurement, $A$ discloses her basis choice.  $B$ discloses the following information: whether he reflected, or measured and, if the latter, whether he got a detection or not.  Note that he does \emph{not} disclose which of the three measurement choices he made - he only discloses that he chose one of them and whether that led to a detection or not.

  \item If $A$ choose the $Z$ basis, and if $B$ choose to measure and did not see a photon, they will use this iteration for their raw key.  Namely, $A$ will use her measurement result and $B$ will use the opposite measurement choice he made (i.e., if he chose \texttt{Measure$-j$} and did not see the photon, his raw key bit will be $1-j$).

  \item A suitable subset of all iterations are chosen and all choices and results are disclosed on this subset to determine the error in the channel.
\end{enumerate}

In the same paper, this protocol was proven to be robust.  Note that it never requires $B$ to send newly-prepared qubits.  Instead, the idea is that he makes a partial measurement of only the $\ket{0}$ or the $\ket{1}$ states; if he does not see the photon (which happens, ideally, with probability $1/2$) then he knows it's been projected to the opposite state.  That is, if he uses \texttt{Measure$-j$} and does not see the photon, it should be leaving his lab in the state $\ket{1-j}$).  Then, when $A$ later measures in the $Z$ basis, she should receive the outcome $1-j$.  A version of this protocol was experimentally implemented in \cite{sqkd-exp1}.  Interestingly, it was shown in \cite{sqkd-mirror-simple} that if this protocol is simplified to remove the \texttt{Measure-All} operation, the protocol is insecure.

An alternative SQKD protocol for practical implementations was presented in \cite{sqkd-reflect-prac}, based off of the Reflection-Based SQKD protocol from \cite{krawec2014restricted}.  Again, the protocol was constructed so that $B$ never had to prepare fresh photons.  Security was shown only against a few practical attacks, namely an unambiguous state discrimination attack similar to the one used against B92 \cite{QKD-B92-USD}, and a multi-photon attack assuming imperfect devices.

Finally, a mediated SQKD protocol was developed in \cite{med-prac} where a fully quantum server prepares and later measures photons.  The two classical users need only to choose $\R$ or to \texttt{Measure}.  They do not need to prepare photons; in fact, they also do not need to measure in a particular basis - they simply need to ``look'' at their portion of the quantum channel thus showing key distribution is possible with very minimal resources.  In the same paper, a complete security proof against collective attacks in the finite-key setting was derived, including device imperfections and assuming an adversarial server (which may even prepare multi-photon states maliciously).  Finally, an experimental demonstration of this protocol was performed and the key-rate computed using these experimental observations thus showing its practicality and the potential for practical semi-quantum communication.

The remaining major challenges, namely the need to switch rapidly from $\MR$ (or some equivalent operation) and $\R$ and, finally device imperfections, remain a challenge.  The latter (e.g., dark counts, loss, and detector efficiency) affects all QKD (semi-quantum and otherwise) work and these should be accounted for in proofs.  Indeed, in the semi-quantum case, they have been accounted for in the papers we consider in this section.  For the first, perhaps new protocols can be developed which do not require rapid switching, or some alternative mechanism for switching can be developed.  For instance, in \cite{passive-switch}, an alternative switching technique using only passive optics was proposed.  Perhaps also the mediated model presents a solution to both: we now know practical mediated SQKD protocols can be built consisting of an (untrusted) quantum server and several classical users.  In the future, as the technology becomes more capable, one can envision only requiring a few commercial centers needing to purchase this expensive technology while end-users need only basic, perhaps poorly performing (e.g., detectors with low efficiency), quantum devices.  Moving forward, when investigating practical semi-quantum communication, these are issues to keep in mind, and device imperfections, along with solutions for mitigating them (perhaps through the use of central servers with good devices, thus allowing end users to have less efficient devices) is an important area of investigation.

Considering the relative ease with which fully-quantum, one-way protocols (such as BB84) may be implemented, it is important to consider how the semi-quantum model may fully contribute beneficially to practical quantum communication.  It seems that several avenues are potentially available: (1) if devices ``break down'' theoretical work within the semi-quantum communication model show that secure communication may still be possible with fewer resources, perhaps by changing the classical post processing; (2) the techniques developed to study these ``limited resource'' protocols, can translate to novel practical insights creating more efficient fully-quantum systems; (3) one may ``offload'' expensive devices to centralized, but untrusted, servers, leaving end-users with cheap, potentially poorly performing, quantum devices yet still attain optimistic security results.  \emph{Research in semi-quantum communication can lead to insights and breakthroughs in these, and other, areas!}

\section{Closing Remarks and Future Directions}

Semi-quantum cryptography and communication was originally introduced to study the theoretical question: how quantum must a protocol be to gain an advantage over its classical counterpart.  This has led to developments in quantum key distribution (namely, semi-quantum key distribution) showing that it is possible to establish a shared secret key, secure against a computationally unbounded adversary, when users have fewer theoretical quantum capabilities.  Namely, even when one user is restricted to ``classical'' operations.  Beyond this, these protocols have even been shown to be comparable in noise tolerance to fully-quantum protocols, at least in ideal perfect qubit channels.  Furthermore, exciting possibilities exist involving semi-quantum users with weak quantum abilities, being able to perform certain cryptographic tasks using the help of strong, but untrusted (and potentially adversarial) servers.  Moving beyond key distribution, the semi-quantum model of communication has been applied to other cryptographic primitives including secret sharing, state comparison, and secure direct communication, to list a few.

This paper has surveyed the history of semi-quantum cryptography and the current state of the art.  We have also discussed recent research in practical, experimental, semi-quantum communication showing that this is a potentially viable model.  There still remains many interesting theoretical and experimental problems, only some of which we have highlighted throughout this review.  On the theory side, it is interesting to see how far one can go in reducing resource requirements and how this affects security.  On the experimental side, it is interesting to see what systems can be built and how.

We believe that research in semi-quantum cryptography can offer great insight into other fields of quantum information science.  The tools and techniques that have been, and are being, developed to construct and analyze semi-quantum protocols can be applied to fully-quantum protocols.  We can gain insight into when security is possible and how to compensate for limited quantum capabilities - all of which are important problems for standard, fully-quantum, systems.  It also provides insight into the great importance of quantum and classical information processing - indeed, many results in semi-quantum cryptography have shown how some lack of a quantum resource may be compensated for by using purely classical means.  There are still many exciting questions and research directions in this area which may shed light on fundamental issues within quantum and classical information science and cryptography.


\begin{thebibliography}{100}

\bibitem{katz-crypto}
Jonathan Katz and Yehuda Lindell.
\newblock {\em Introduction to modern cryptography}.
\newblock Chapman and Hall/CRC, 2014.

\bibitem{stallings2007network}
William Stallings.
\newblock {\em Network security essentials: applications and standards}.
\newblock Pearson Education India, 2007.

\bibitem{QKD-BB84}
Charles~H Bennett and Gilles Brassard.
\newblock Quantum cryptography: Public key distribution and coin tossing.
\newblock In {\em Proceedings of IEEE International Conference on Computers,
  Systems and Signal Processing}, volume 175. New York, 1984.

\bibitem{QKD-E91}
Artur~K Ekert.
\newblock Quantum cryptography based on bell’s theorem.
\newblock {\em Physical review letters}, 67(6):661, 1991.

\bibitem{QKD-BB84-rate1}
Peter~W. Shor and John Preskill.
\newblock Simple proof of security of the bb84 quantum key distribution
  protocol.
\newblock {\em Phys. Rev. Lett.}, 85:441--444, Jul 2000.

\bibitem{QKD-renner-keyrate}
Renato Renner, Nicolas Gisin, and Barbara Kraus.
\newblock Information-theoretic security proof for quantum-key-distribution
  protocols.
\newblock {\em Phys. Rev. A}, 72:012332, Jul 2005.

\bibitem{SQKD-first}
Michel Boyer, Dan Kenigsberg, and Tal Mor.
\newblock Quantum key distribution with classical bob.
\newblock {\em Phys. Rev. Lett.}, 99:140501, Oct 2007.

\bibitem{qkd-survey}
Valerio Scarani, Helle Bechmann-Pasquinucci, Nicolas~J. Cerf, Miloslav
  Du\ifmmode~\check{s}\else \v{s}\fi{}ek, Norbert L\"utkenhaus, and Momtchil
  Peev.
\newblock The security of practical quantum key distribution.
\newblock {\em Rev. Mod. Phys.}, 81:1301--1350, Sep 2009.

\bibitem{qkd-survey2}
Akshata Shenoy-Hejamadi, Anirban Pathak, and Srikanth Radhakrishna.
\newblock Quantum cryptography: Key distribution and beyond.
\newblock {\em Quanta}, 6(1):1--47, 2017.

\bibitem{qkd-survey3}
Mohsen Razavi, Anthony Leverrier, Xiongfeng Ma, Bing Qi, and Zhiliang Yuan.
\newblock Quantum key distribution and beyond: introduction.
\newblock {\em J. Opt. Soc. Am. B}, 36(3):QKD1--QKD2, Mar 2019.

\bibitem{qkd-survey4}
S~Pirandola, UL~Andersen, L~Banchi, M~Berta, D~Bunandar, R~Colbeck, D~Englund,
  T~Gehring, C~Lupo, C~Ottaviani, et~al.
\newblock Advances in quantum cryptography.
\newblock {\em arXiv preprint arXiv:1906.01645}, 2019.

\bibitem{sqkd-second}
Michel Boyer, Ran Gelles, Dan Kenigsberg, and Tal Mor.
\newblock Semiquantum key distribution.
\newblock {\em Phys. Rev. A}, 79:032341, Mar 2009.

\bibitem{ent1}
Wang Jian, Zhang Sheng, Zhang Quan, and Tang Chao-Jing.
\newblock Semiquantum key distribution using entangled states.
\newblock {\em Chinese Physics Letters}, 28(10):100301, 2011.

\bibitem{ent2}
Zhiwei Sun, Ruigang Du, and Dongyang Long.
\newblock Semi-quantum key distribution protocol using bell state.
\newblock {\em arXiv preprint arXiv:1106.2910}, 2011.

\bibitem{SQKD-less-than-4}
Xiangfu Zou, Daowen Qiu, Lvzhou Li, Lihua Wu, and Lvjun Li.
\newblock Semiquantum-key distribution using less than four quantum states.
\newblock {\em Phys. Rev. A}, 79:052312, May 2009.

\bibitem{krawec2014restricted}
Walter~O Krawec.
\newblock Restricted attacks on semi-quantum key distribution protocols.
\newblock {\em Quantum Information Processing}, 13(11):2417--2436, 2014.

\bibitem{QKD-BB84-three-state}
Chi-Hang~Fred Fung and Hoi-Kwong Lo.
\newblock Security proof of a three-state quantum-key-distribution protocol
  without rotational symmetry.
\newblock {\em Phys. Rev. A}, 74:042342, Oct 2006.

\bibitem{QKD-BB84-three-state-v2}
Cyril Branciard, Nicolas Gisin, Norbert Lutkenhaus, and Valerio Scarani.
\newblock Zero-error attacks and detection statistics in the coherent one-way
  protocol for quantum cryptography.
\newblock {\em Quantum Information \& Computation}, 7(7):639--664, 2007.

\bibitem{QKD-B92-extended}
Marco Lucamarini, Giovanni Di~Giuseppe, and Kiyoshi Tamaki.
\newblock Robust unconditionally secure quantum key distribution with two
  nonorthogonal and uninformative states.
\newblock {\em Physical Review A}, 80(3):032327, 2009.

\bibitem{sqkd-single-state-b92}
Wei Zhang and Daowen Qiu.
\newblock A single-state semi-quantum key distribution protocol and its
  security proof.
\newblock {\em arXiv preprint arXiv:1612.03087}, 2016.

\bibitem{cl-A}
Hua Lu and Qing-Yu Cai.
\newblock Quantum key distribution with classical alice.
\newblock {\em International Journal of Quantum Information}, 6(06):1195--1202,
  2008.

\bibitem{refresh}
Zhi-Wei Sun, Rui-Gang Du, and Dong-Yang Long.
\newblock Quantum key distribution with limited classical bob.
\newblock {\em International Journal of Quantum Information}, 11(01):1350005,
  2013.

\bibitem{double-cnot}
Po-Hua Lin, Tzonelih Hwang, and Chia-Wei Tsai.
\newblock Double cnot attack on “quantum key distribution with limited
  classical bob”.
\newblock {\em International Journal of Quantum Information}, 17(02):1975001,
  2019.

\bibitem{sqkd-limited-measure}
Walter~O Krawec and Eric~P Geiss.
\newblock Semi-quantum key distribution with limited measurement capabilities.
\newblock In {\em 2018 International Symposium on Information Theory and Its
  Applications (ISITA)}, pages 462--466. IEEE, 2018.

\bibitem{sqkd-classical-quantum}
Allison Gagliano, Walter~O Krawec, and Hasan Iqbal.
\newblock From classical to semi-quantum secure communication.
\newblock In {\em 2019 IEEE International Symposium on Information Theory
  (ISIT)}, pages 1707--1711. IEEE, 2019.

\bibitem{no-measure}
Xiangfu Zou, Daowen Qiu, Shengyu Zhang, and Paulo Mateus.
\newblock Semiquantum key distribution without invoking the classical party’s
  measurement capability.
\newblock {\em Quantum Information Processing}, 14(8):2981--2996, 2015.

\bibitem{no-measure-2}
Qin Li, Wai~Hong Chan, and Shengyu Zhang.
\newblock Semiquantum key distribution with secure delegated quantum
  computation.
\newblock {\em Scientific reports}, 6:19898, 2016.

\bibitem{sqkd-no-auth-1}
Kun-Fei Yu, Chun-Wei Yang, Ci-Hong Liao, and Tzonelih Hwang.
\newblock Authenticated semi-quantum key distribution protocol using bell
  states.
\newblock {\em Quantum Information Processing}, 13(6):1457--1465, 2014.

\bibitem{sqkd-no-auth-2}
Chuan-Ming Li, Kun-Fei Yu, Shih-Hung Kao, and Tzonelih Hwang.
\newblock Authenticated semi-quantum key distributions without classical
  channel.
\newblock {\em Quantum Information Processing}, 15(7):2881--2893, 2016.

\bibitem{sqkd-no-auth-atk}
A~Meslouhi and Yassine Hassouni.
\newblock Cryptanalysis on authenticated semi-quantum key distribution protocol
  using bell states.
\newblock {\em Quantum Information Processing}, 16(1):18, 2017.

\bibitem{sqkd-high-noise}
Omar Amer and Walter~O Krawec.
\newblock Semiquantum key distribution with high quantum noise tolerance.
\newblock {\em Physical Review A}, 100(2):022319, 2019.

\bibitem{sqkd-eff-0}
Wei Liu and Huaijun Zhou.
\newblock A new semi-quantum key distribution protocol with high efficiency.
\newblock In {\em 2018 IEEE 3rd Advanced Information Technology, Electronic and
  Automation Control Conference (IAEAC)}, pages 2424--2427. IEEE, 2018.

\bibitem{sqkd-eff}
Ming-Ming Wang, Lin-Ming Gong, and Lian-He Shao.
\newblock Efficient semiquantum key distribution without entanglement.
\newblock {\em Quantum Information Processing}, 18(9):260, 2019.

\bibitem{QKD-BB84-Modification}
Hoi-Kwong Lo, Hoi-Fung Chau, and M~Ardehali.
\newblock Efficient quantum key distribution scheme and a proof of its
  unconditional security.
\newblock {\em Journal of Cryptology}, 18(2):133--165, 2005.

\bibitem{sqkd-dephase}
Ming-Hui Zhang, Hui-Fang Li, Jin-Ye Peng, and Xiao-Yi Feng.
\newblock Fault-tolerant semiquantum key distribution over a
  collective-dephasing noise channel.
\newblock {\em International Journal of Theoretical Physics}, 56(8):2659--2670,
  2017.

\bibitem{sqkd-dephase-rotate}
Chih-Lun Tsai and Tzonelih Hwang.
\newblock Semi-quantum key distribution robust against combined collective
  noise.
\newblock {\em International Journal of Theoretical Physics},
  57(11):3410--3418, 2018.

\bibitem{sqkd-dephase-rotate-2}
Chia-Wei Tsai and Chun-Wei Yang.
\newblock Cryptanalysis and improvement of the semi-quantum key distribution
  robust against combined collective noise.
\newblock {\em International Journal of Theoretical Physics}, pages 1--7, 2019.

\bibitem{sqkd-qw}
Chrysoula Vlachou, Walter Krawec, Paulo Mateus, Nikola Paunkovi{\'c}, and
  Andr{\'e} Souto.
\newblock Quantum key distribution with quantum walks.
\newblock {\em Quantum Information Processing}, 17(11):288, 2018.

\bibitem{sqkd-high-dim}
Hasan Iqbal and Walter~O Krawec.
\newblock High-dimensional semi-quantum cryptography.
\newblock {\em arXiv preprint arXiv:1907.11340}, 2019.

\bibitem{high-dim0}
H~Bechmann-Pasquinucci and Wolfgang Tittel.
\newblock Quantum cryptography using larger alphabets.
\newblock {\em Physical Review A}, 61(6):062308, 2000.

\bibitem{QKD-high-dim-50}
HF~Chau.
\newblock Quantum key distribution using qudits that each encode one bit of raw
  key.
\newblock {\em Physical Review A}, 92(6):062324, 2015.

\bibitem{high-dim5}
Toshihiko Sasaki, Yoshihisa Yamamoto, and Masato Koashi.
\newblock Practical quantum key distribution protocol without monitoring signal
  disturbance.
\newblock {\em Nature}, 509(7501):475, 2014.

\bibitem{high-dim6}
Zhen-Qiang Yin, Shuang Wang, Wei Chen, Yun-Guang Han, Rong Wang, Guang-Can Guo,
  and Zheng-Fu Han.
\newblock Improved security bound for the round-robin-differential-phase-shift
  quantum key distribution.
\newblock {\em Nature communications}, 9(1):457, 2018.

\bibitem{high-dim7}
Rong Wang, Zhen-Qiang Yin, Chao-han Cui, Shuang Wang, Wei Chen, Guang-Can Guo,
  and Zheng-Fu Han.
\newblock Security proof for single-photon round-robin
  differential-quadrature-phase-shift quantum key distribution.
\newblock {\em Physical Review A}, 98(6):062331, 2018.

\bibitem{QW-intro1}
Julia Kempe.
\newblock Quantum random walks: an introductory overview.
\newblock {\em Contemporary Physics}, 44(4):307--327, 2003.

\bibitem{QW-survey}
Salvador~El{\'\i}as Venegas-Andraca.
\newblock Quantum walks: a comprehensive review.
\newblock {\em Quantum Information Processing}, 11(5):1015--1106, 2012.

\bibitem{sqkd-eavesdropping}
Arpita Maitra and Goutam Paul.
\newblock Eavesdropping in semiquantum key distribution protocol.
\newblock {\em Information Processing Letters}, 113(12):418--422, 2013.

\bibitem{QKD-Winter-Keyrate}
Igor Devetak and Andreas Winter.
\newblock Distillation of secret key and entanglement from quantum states.
\newblock {\em Proceedings of the Royal Society A: Mathematical, Physical and
  Engineering Science}, 461(2053):207--235, 2005.

\bibitem{info-disturbance}
Takayuki Miyadera.
\newblock Relation between information and disturbance in quantum key
  distribution protocol with classical alice.
\newblock {\em International Journal of Quantum Information}, 9(06):1427--1435,
  2011.

\bibitem{BB84-ind1}
Christopher~A Fuchs, Nicolas Gisin, Robert~B Griffiths, Chi-Sheng Niu, and
  Asher Peres.
\newblock Optimal eavesdropping in quantum cryptography. i. information bound
  and optimal strategy.
\newblock {\em Physical Review A}, 56(2):1163, 1997.

\bibitem{krawec-entropic}
Walter~O Krawec.
\newblock Key-rate bound of a semi-quantum protocol using an entropic
  uncertainty relation.
\newblock In {\em 2018 IEEE International Symposium on Information Theory
  (ISIT)}, pages 2669--2673. IEEE, 2018.

\bibitem{krawec2015security}
Walter~O Krawec.
\newblock Security proof of a semi-quantum key distribution protocol.
\newblock In {\em 2015 IEEE International Symposium on Information Theory
  (ISIT)}, pages 686--690. IEEE, 2015.

\bibitem{QC-alicki2004continuity}
Robert Alicki and Mark Fannes.
\newblock Continuity of quantum conditional information.
\newblock {\em Journal of Physics A: Mathematical and General}, 37(5):L55,
  2004.

\bibitem{QC-fannes-audenaert-inequality}
Koenraad~MR Audenaert.
\newblock A sharp continuity estimate for the von neumann entropy.
\newblock {\em Journal of Physics A: Mathematical and Theoretical},
  40(28):8127, 2007.

\bibitem{QC-winter2016tight}
Andreas Winter.
\newblock Tight uniform continuity bounds for quantum entropies: conditional
  entropy, relative entropy distance and energy constraints.
\newblock {\em Communications in Mathematical Physics}, 347(1):291--313, 2016.

\bibitem{QKD-uncertainty}
Mario Berta, Matthias Christandl, Roger Colbeck, Joseph~M Renes, and Renato
  Renner.
\newblock The uncertainty principle in the presence of quantum memory.
\newblock {\em Nature Physics}, 6(9):659--662, 2010.

\bibitem{uncertainty-survey1}
Patrick~J. Coles, Mario Berta, Marco Tomamichel, and Stephanie Wehner.
\newblock Entropic uncertainty relations and their applications.
\newblock {\em Rev. Mod. Phys.}, 89:015002, Feb 2017.

\bibitem{uncertainty-survey-2}
Iwo Bialynicki-Birula and {\L}ukasz Rudnicki.
\newblock Entropic uncertainty relations in quantum physics.
\newblock In {\em Statistical Complexity}, pages 1--34. Springer, 2011.

\bibitem{uncertainty-survey-3}
Stephanie Wehner and Andreas Winter.
\newblock Entropic uncertainty relations—a survey.
\newblock {\em New Journal of Physics}, 12(2):025009, 2010.

\bibitem{QKD-Tom-Krawec-Arbitrary}
Walter~O. Krawec.
\newblock Quantum key distribution with mismatched measurements over arbitrary
  channels.
\newblock {\em Quantum Information and Computation}, 17(3 and 4):209--241,
  2017.

\bibitem{QKD-TwoWaySecure}
Normand~J Beaudry, Marco Lucamarini, Stefano Mancini, and Renato Renner.
\newblock Security of two-way quantum key distribution.
\newblock {\em Physical Review A}, 88(6):062302, 2013.

\bibitem{lucamarini2014quantum}
Marco Lucamarini and Stefano Mancini.
\newblock Quantum key distribution using a two-way quantum channel.
\newblock {\em Theoretical Computer Science}, 560:46--61, 2014.

\bibitem{SQKD-Krawec-dissertation}
Walter~O Krawec.
\newblock {\em Semi-Quantum Key Distribution: Protocols, Security Analysis, and
  New Models}.
\newblock PhD thesis, Stevens Institute of Technology, May 2015.

\bibitem{krawec-reflect-new}
Walter~O Krawec.
\newblock Security of a semi-quantum protocol where reflections contribute to
  the secret key.
\newblock {\em Quantum Information Processing}, 15(5):2067--2090, 2016.

\bibitem{QKD-Tom-First}
Stephen~M Barnett, Bruno Huttner, and Simon~JD Phoenix.
\newblock Eavesdropping strategies and rejected-data protocols in quantum
  cryptography.
\newblock {\em Journal of Modern Optics}, 40(12):2501--2513, 1993.

\bibitem{QKD-Tom-KeyRateIncrease}
Shun Watanabe, Ryutaroh Matsumoto, and Tomohiko Uyematsu.
\newblock Tomography increases key rates of quantum-key-distribution protocols.
\newblock {\em Physical Review A}, 78(4):042316, 2008.

\bibitem{QKD-Tom-KeyRateMismatchedDistill}
Ryutaroh Matsumoto and Shun Watanabe.
\newblock Key rate available from mismatched measurements in the bb84 protocol
  and the uncertainty principle.
\newblock {\em IEICE Transactions on Fundamentals of Electronics,
  Communications and Computer Sciences}, 91(10):2870--2873, 2008.

\bibitem{QKD-Tom-BB84NarrowAngle}
Ryutaroh Matsumoto and Shun Watanabe.
\newblock Narrow basis angle doubles secret key in the bb84 protocol.
\newblock {\em Journal of Physics A: Mathematical and Theoretical},
  43(14):145302, 2010.

\bibitem{QKD-Tom-threestate1}
Kiyoshi Tamaki, Marcos Curty, Go~Kato, Hoi-Kwong Lo, and Koji Azuma.
\newblock Loss-tolerant quantum cryptography with imperfect sources.
\newblock {\em Physical Review A}, 90(5):052314, 2014.

\bibitem{QKD-Tom-threestate-Krawec}
Walter~O. Krawec.
\newblock Asymptotic analysis of a three state quantum cryptographic protocol.
\newblock In {\em {IEEE} International Symposium on Information Theory, {ISIT}
  2016, Barcelona, July 10-15, 2016}, pages 2489--2493, 2016.

\bibitem{zhang2018security}
Wei Zhang, Daowen Qiu, and Paulo Mateus.
\newblock Security of a single-state semi-quantum key distribution protocol.
\newblock {\em Quantum Information Processing}, 17(6):135, 2018.

\bibitem{multi1}
Zhang Xian-Zhou, Gong Wei-Gui, Tan Yong-Gang, Ren Zhen-Zhong, and Guo
  Xiao-Tian.
\newblock Quantum key distribution series network protocol with m-classical
  bobs.
\newblock {\em Chinese Physics B}, 18(6):2143, 2009.

\bibitem{multi2}
Kong-Ni Zhu, Nan-Run Zhou, Yun-Qian Wang, and Xiao-Jun Wen.
\newblock Semi-quantum key distribution protocols with ghz states.
\newblock {\em International Journal of Theoretical Physics},
  57(12):3621--3631, 2018.

\bibitem{multi3}
Nan-Run Zhou, Kong-Ni Zhu, and Xiang-Fu Zou.
\newblock Multi-party semi-quantum key distribution protocol with four-particle
  cluster states.
\newblock {\em Annalen der Physik}, page 1800520, 2019.

\bibitem{cluster1}
Robert Raussendorf and Hans~J Briegel.
\newblock A one-way quantum computer.
\newblock {\em Physical Review Letters}, 86(22):5188, 2001.

\bibitem{sqkd-med-first}
Walter~O Krawec.
\newblock Mediated semiquantum key distribution.
\newblock {\em Physical Review A}, 91(3):032323, 2015.

\bibitem{sqkd-med-improved}
Walter~O Krawec.
\newblock An improved asymptotic key rate bound for a mediated semi-quantum key
  distribution protocol.
\newblock {\em Quantum Information and Computation}, 16(9 and 10):813--834,
  2016.

\bibitem{mm-sqkd}
Walter~O Krawec.
\newblock Multi-mediated semi-quantum key distribution.
\newblock {\em To appear: 2019 IEEE Globecom Workshops (GC Wkshps)}, 2019.

\bibitem{sqkd-med-no-measure}
Zhi-Rou Liu and Tzonelih Hwang.
\newblock Mediated semi-quantum key distribution without invoking quantum
  measurement.
\newblock {\em Annalen der Physik}, 530(4):1700206, 2018.

\bibitem{med-single-photon}
Po-Hua Lin, Chia-Wei Tsai, and Tzonelih Hwang.
\newblock Mediated semi-quantum key distribution using single photons.
\newblock {\em Annalen der Physik}, page 1800347, 2019.

\bibitem{med-light}
Chia-Wei Tsai and Chun-Wei Yang.
\newblock Lightweight mediated semi-quantum key distribution protocol with a
  dishonest third party based on bell states.
\newblock {\em arXiv preprint arXiv:1909.02788}, 2019.

\bibitem{med-light-2}
Chia-Wei Tsai, Chun-Wei Yang, and Narn-Yih Lee.
\newblock Lightweight mediated semi-quantum key distribution protocol.
\newblock {\em Modern Physics Letters A}, page 1950281, 2019.

\bibitem{med-prac}
Francesco Massa, Preeti Yadav, Amir Moqanaki, Walter~O Krawec, Paulo Mateus,
  Nikola Paunkovi{\'c}, Andr{\'e} Souto, and Philip Walther.
\newblock Experimental quantum cryptography with classical users.
\newblock {\em arXiv preprint arXiv:1908.01780}, 2019.

\bibitem{secret-orig}
Adi Shamir.
\newblock How to share a secret.
\newblock {\em Communications of the ACM}, 22(11):612--613, 1979.

\bibitem{secret-survey}
Amos Beimel.
\newblock Secret-sharing schemes: a survey.
\newblock In {\em International Conference on Coding and Cryptology}, pages
  11--46. Springer, 2011.

\bibitem{secret-quantum-first}
Mark Hillery, Vladim{\'\i}r Bu{\v{z}}ek, and Andr{\'e} Berthiaume.
\newblock Quantum secret sharing.
\newblock {\em Physical Review A}, 59(3):1829, 1999.

\bibitem{secret-quantum-second}
Anders Karlsson, Masato Koashi, and Nobuyuki Imoto.
\newblock Quantum entanglement for secret sharing and secret splitting.
\newblock {\em Physical Review A}, 59(1):162, 1999.

\bibitem{secret-quantum-third}
Daniel Gottesman.
\newblock Theory of quantum secret sharing.
\newblock {\em Physical Review A}, 61(4):042311, 2000.

\bibitem{sqkd-secret-first}
Qin Li, Wai~Hong Chan, and Dong-Yang Long.
\newblock Semiquantum secret sharing using entangled states.
\newblock {\em Physical Review A}, 82(2):022303, 2010.

\bibitem{sqkd-secret-second}
Jian Wang, Sheng Zhang, Quan Zhang, and Chao-Jing Tang.
\newblock Semiquantum secret sharing using two-particle entangled state.
\newblock {\em International Journal of Quantum Information}, 10(05):1250050,
  2012.

\bibitem{sqkd-secret-multiparty}
Lvzhou Li, Daowen Qiu, and Paulo Mateus.
\newblock Quantum secret sharing with classical bobs.
\newblock {\em Journal of Physics A: Mathematical and Theoretical},
  46(4):045304, 2013.

\bibitem{sqkd-secret-message}
Chen Xie, Lvzhou Li, and Daowen Qiu.
\newblock A novel semi-quantum secret sharing scheme of specific bits.
\newblock {\em International Journal of Theoretical Physics},
  54(10):3819--3824, 2015.

\bibitem{sqkd-secret-message-atk}
Aihan Yin and Fangbo Fu.
\newblock Eavesdropping on semi-quantum secret sharing scheme of specific bits.
\newblock {\em International Journal of Theoretical Physics}, 55(9):4027--4035,
  2016.

\bibitem{sqkd-secret-message-atk-2}
Xiang Gao, Shibin Zhang, and Yan Chang.
\newblock Cryptanalysis and improvement of the semi-quantum secret sharing
  protocol.
\newblock {\em International Journal of Theoretical Physics}, 56(8):2512--2520,
  2017.

\bibitem{sqkd-secret-limited}
Yi~Xiang, Jun Liu, Ming-qiang Bai, Xue Yang, and Zhi-wen Mo.
\newblock Limited resource semi-quantum secret sharing based on multi-level
  systems.
\newblock {\em International Journal of Theoretical Physics}, 58(9):2883--2892,
  2019.

\bibitem{sqkd-secret-circle}
Chong-Qiang Ye and Tian-Yu Ye.
\newblock Circular semi-quantum secret sharing using single particles.
\newblock {\em Communications in Theoretical Physics}, 70(6):661, 2018.

\bibitem{sqkd-secret-limited-2}
Zhulin Li, Qin Li, Chengdong Liu, Yu~Peng, Wai~Hong Chan, and Lvzhou Li.
\newblock Limited resource semiquantum secret sharing.
\newblock {\em Quantum Information Processing}, 17(10):285, 2018.

\bibitem{high-dim-secret-share}
Ye~Chong-Qiang, Ye~Tian-Yu, He~De, and Gan Zhi-Gang.
\newblock Multiparty semi-quantum secret sharing with d-level single-particle
  states.
\newblock {\em International Journal of Theoretical Physics}, pages 1--18,
  2019.

\bibitem{secret-W-state}
Chia-Wei Tsai, Chun-Wei Yang, and Narn-Yih Lee.
\newblock Semi-quantum secret sharing protocol using w-state.
\newblock {\em Modern Physics Letters A}, 34(27):1950213, 2019.

\bibitem{sqkd-secret-convert}
Kun-Fei Yu, Jun Gu, Tzonelih Hwang, and Prosanta Gope.
\newblock Multi-party semi-quantum key distribution-convertible multi-party
  semi-quantum secret sharing.
\newblock {\em Quantum Information Processing}, 16(8):194, 2017.

\bibitem{sqkd-secret-bell}
Aihan Yin, Zefan Wang, and Fangbo Fu.
\newblock A novel semi-quantum secret sharing scheme based on bell states.
\newblock {\em Modern Physics Letters B}, 31(13):1750150, 2017.

\bibitem{sqkd-secret-bell-atk}
Gan Gao, Yue Wang, and Dong Wang.
\newblock Cryptanalysis of a semi-quantum secret sharing scheme based on bell
  states.
\newblock {\em Modern Physics Letters B}, 32(09):1850117, 2018.

\bibitem{sqkd-secret-bell-2}
Gan Gao, Yue Wang, and Dong Wang.
\newblock Multiparty semiquantum secret sharing based on rearranging orders of
  qubits.
\newblock {\em Modern Physics Letters B}, 30(10):1650130, 2016.

\bibitem{sqkd-secret-arb}
Ai~Han Yin and Yan Tong.
\newblock A novel semi-quantum secret sharing scheme using entangled states.
\newblock {\em Modern Physics Letters B}, 32(22):1850256, 2018.

\bibitem{sqkd-secret-arb-atk}
Qijian He, Wei Yang, Bingren Chen, and Liusheng Huang.
\newblock Cryptanalysis and improvement of the novel semi-quantum secret
  sharing scheme using entangled states.
\newblock {\em Modern Physics Letters B}, 33(04):1950045, 2019.

\bibitem{sqss-scale}
Gang Cao, Chen Chen, and Min Jiang.
\newblock A scalable and flexible multi-user semi-quantum secret sharing.
\newblock In {\em Proceedings of the 2nd International Conference on
  Telecommunications and Communication Engineering}, pages 28--32. ACM, 2018.

\bibitem{sqss-quantum}
Yi-you Nie, Yuan-hua Li, and Zi-sheng Wang.
\newblock Semi-quantum information splitting using ghz-type states.
\newblock {\em Quantum information processing}, 12(1):437--448, 2013.

\bibitem{fully-sdc0}
Gui-Lu Long and Xiao-Shu Liu.
\newblock Theoretically efficient high-capacity quantum-key-distribution
  scheme.
\newblock {\em Physical Review A}, 65(3):032302, 2002.

\bibitem{fully-sdc00}
Kim Bostr{\"o}m and Timo Felbinger.
\newblock Deterministic secure direct communication using entanglement.
\newblock {\em Physical Review Letters}, 89(18):187902, 2002.

\bibitem{fully-sdc1}
Fu-Guo Deng, Gui~Lu Long, and Xiao-Shu Liu.
\newblock Two-step quantum direct communication protocol using the
  einstein-podolsky-rosen pair block.
\newblock {\em Physical Review A}, 68(4):042317, 2003.

\bibitem{fully-sdc2}
G.~{Long}.
\newblock Quantum secure direct communication: Principles, current status,
  perspectives.
\newblock In {\em 2017 IEEE 85th Vehicular Technology Conference (VTC Spring)},
  pages 1--5, June 2017.

\bibitem{fully-sdc3}
Fu-Guo Deng and Gui~Lu Long.
\newblock Secure direct communication with a quantum one-time pad.
\newblock {\em Physical Review A}, 69(5):052319, 2004.

\bibitem{sqdc-first}
XiangFu Zou and DaoWen Qiu.
\newblock Three-step semiquantum secure direct communication protocol.
\newblock {\em Science China Physics, Mechanics \& Astronomy},
  57(9):1696--1702, 2014.

\bibitem{sqdc-attack}
Jun Gu, Po-hua Lin, and Tzonelih Hwang.
\newblock Double c-not attack and counterattack on ‘three-step semi-quantum
  secure direct communication protocol’.
\newblock {\em Quantum Information Processing}, 17(7):182, 2018.

\bibitem{sqdc-2}
Chen Xie, Lvzhou Li, Haozhen Situ, and Jianhao He.
\newblock Semi-quantum secure direct communication scheme based on bell states.
\newblock {\em International Journal of Theoretical Physics}, 57(6):1881--1887,
  2018.

\bibitem{sqdc-bell}
Ming-Hui Zhang, Hui-Fang Li, Zhao-Qiang Xia, Xiao-Yi Feng, and Jin-Ye Peng.
\newblock Semiquantum secure direct communication using epr pairs.
\newblock {\em Quantum Information Processing}, 16(5):117, 2017.

\bibitem{sqdc-bell-2}
LiLi Yan, YuHua Sun, Yan Chang, ShiBin Zhang, GuoGen Wan, and ZhiWei Sheng.
\newblock Semi-quantum protocol for deterministic secure quantum communication
  using bell states.
\newblock {\em Quantum Information Processing}, 17(11):315, 2018.

\bibitem{sqdc-two-eff}
Yuhua Sun, Lili Yan, Yan Chang, Shibin Zhang, Tingting Shao, and Yan Zhang.
\newblock Two semi-quantum secure direct communication protocols based on bell
  states.
\newblock {\em Modern Physics Letters A}, 34(01):1950004, 2019.

\bibitem{sqdc-auth}
Yi-Ping Luo and Tzonelih Hwang.
\newblock Authenticated semi-quantum direct communication protocols using bell
  states.
\newblock {\em Quantum Information Processing}, 15(2):947--958, 2016.

\bibitem{sqdc-auth-atk}
Saleh Almousa and Michel Barbeau.
\newblock Delay and reflection attacks in authenticated semi-quantum direct
  communications.
\newblock In {\em 2016 IEEE Globecom Workshops (GC Wkshps)}, pages 1--7. IEEE,
  2016.

\bibitem{sqdc-auth-eco}
Haoye Lu, Michel Barbeau, and Amiya Nayak.
\newblock Economic no-key semi-quantum direct communication protocol.
\newblock In {\em 2017 IEEE Globecom Workshops (GC Wkshps)}, pages 1--7. IEEE,
  2017.

\bibitem{sqdc-auth-eco-2}
Haoye Lu, Michel Barbeau, and Amiya Nayak.
\newblock Keyless semi-quantum point-to-point communication protocol with low
  resource requirements.
\newblock {\em Scientific reports}, 9(1):64, 2019.

\bibitem{sqkd-sdc-auth-single}
Ming-Ming Wang, Jun-Li Liu, and Lin-Ming Gong.
\newblock Semiquantum secure direct communication with authentication based on
  single-photons.
\newblock {\em International Journal of Quantum Information}, page 1950024,
  2019.

\bibitem{sqdc-auth-two}
Zheng Tao, Yan Chang, Shibin Zhang, Jinqiao Dai, and Xueyang Li.
\newblock Two semi-quantum direct communication protocols with mutual
  authentication based on bell states.
\newblock {\em International Journal of Theoretical Physics}, pages 1--8, 2019.

\bibitem{qkd-qd-0}
Zhan-Jun Zhang and Zhong-Xiao Man.
\newblock Secure direct bidirectional communication protocol using the
  einstein-podolsky-rosen pair block.
\newblock {\em arXiv preprint quant-ph/0403215}, 2004.

\bibitem{qkd-qd-1}
Ba~An Nguyen.
\newblock Quantum dialogue.
\newblock {\em Physics Letters A}, 328(1):6--10, 2004.

\bibitem{sqdc-qd-1}
Chitra Shukla, Kishore Thapliyal, and Anirban Pathak.
\newblock Semi-quantum communication: protocols for key agreement, controlled
  secure direct communication and dialogue.
\newblock {\em Quantum Information Processing}, 16(12):295, 2017.

\bibitem{sqdc-qd-2}
Tian-Yu Ye and Chong-Qiang Ye.
\newblock Semi-quantum dialogue based on single photons.
\newblock {\em International Journal of Theoretical Physics}, 57(5):1440--1454,
  2018.

\bibitem{sqdc-qd-auth}
Lin Liu, Min Xiao, and Xiuli Song.
\newblock Authenticated semiquantum dialogue with secure delegated quantum
  computation over a collective noise channel.
\newblock {\em Quantum Information Processing}, 17(12):342, 2018.

\bibitem{QKA-1}
Nanrun Zhou, Guihua Zeng, and Jin Xiong.
\newblock Quantum key agreement protocol.
\newblock {\em Electronics Letters}, 40(18):1149--1150, 2004.

\bibitem{QKA-2}
Song-Kong Chong and Tzonelih Hwang.
\newblock Quantum key agreement protocol based on bb84.
\newblock {\em Optics Communications}, 283(6):1192--1195, 2010.

\bibitem{QKA-3}
Chitra Shukla, Nasir Alam, and Anirban Pathak.
\newblock Protocols of quantum key agreement solely using bell states and bell
  measurement.
\newblock {\em Quantum information processing}, 13(11):2391--2405, 2014.

\bibitem{sqkd-qka-0}
Wen-Jie Liu, Zhen-Yu Chen, Sai Ji, Hai-Bin Wang, and Jun Zhang.
\newblock Multi-party semi-quantum key agreement with delegating quantum
  computation.
\newblock {\em International Journal of Theoretical Physics},
  56(10):3164--3174, 2017.

\bibitem{sqkd-qka-1}
Li~Li Yan, Shi~Bin Zhang, Yan Chang, Zhi~Wei Sheng, and Fan Yang.
\newblock Mutual semi-quantum key agreement protocol using bell states.
\newblock {\em Modern Physics Letters A}, page 1950294, 2019.

\bibitem{sqkd-qka-2}
Lili Yan, Shibin Zhang, Yan Chang, Zhiwei Sheng, and Yuhua Sun.
\newblock Semi-quantum key agreement and private comparison protocols using
  bell states.
\newblock {\em International Journal of Theoretical Physics}, pages 1--11,
  2019.

\bibitem{smc1}
Andrew~C Yao.
\newblock Protocols for secure computations.
\newblock In {\em 23rd annual symposium on foundations of computer science
  (sfcs 1982)}, pages 160--164. IEEE, 1982.

\bibitem{smc2}
Yehida Lindell.
\newblock Secure multiparty computation for privacy preserving data mining.
\newblock In {\em Encyclopedia of Data Warehousing and Mining}, pages
  1005--1009. IGI Global, 2005.

\bibitem{qpc1}
Xiu-Bo Chen, Gang Xu, Xin-Xin Niu, Qiao-Yan Wen, and Yi-Xian Yang.
\newblock An efficient protocol for the private comparison of equal information
  based on the triplet entangled state and single-particle measurement.
\newblock {\em Optics communications}, 283(7):1561--1565, 2010.

\bibitem{qpc2}
Wen Liu, Yong-Bin Wang, and Zheng-Tao Jiang.
\newblock An efficient protocol for the quantum private comparison of equality
  with w state.
\newblock {\em Optics Communications}, 284(12):3160--3163, 2011.

\bibitem{qpc3}
Wen Liu, Yong-Bin Wang, Zheng-Tao Jiang, and Yi-Zhen Cao.
\newblock A protocol for the quantum private comparison of equality with
  $\chi$-type state.
\newblock {\em International Journal of Theoretical Physics}, 51(1):69--77,
  2012.

\bibitem{qpc-survey}
Wenjie Liu, Chao Liu, Haibin Wang, and Tingting Jia.
\newblock Quantum private comparison: a review.
\newblock {\em IETE Technical Review}, 30(5):439--445, 2013.

\bibitem{qpc-imp}
Hoi-Kwong Lo.
\newblock Insecurity of quantum secure computations.
\newblock {\em Physical Review A}, 56(2):1154, 1997.

\bibitem{sqpc-first}
Kishore Thapliyal, Rishi~Dutt Sharma, and Anirban Pathak.
\newblock Orthogonal-state-based and semi-quantum protocols for quantum private
  comparison in noisy environment.
\newblock {\em International Journal of Quantum Information}, 16(05):1850047,
  2018.

\bibitem{sqpc-first-2}
Wen-Han Chou, Tzonelih Hwang, and Jun Gu.
\newblock Semi-quantum private comparison protocol under an almost-dishonest
  third party.
\newblock {\em arXiv preprint arXiv:1607.07961}, 2016.

\bibitem{sqpc-single}
Lang Yan-Feng.
\newblock Semi-quantum private comparison using single photons.
\newblock {\em International Journal of Theoretical Physics},
  57(10):3048--3055, 2018.

\bibitem{sqpc-single-2}
Po-Hua Lin, Tzonelih Hwang, and Chia-Wei Tsai.
\newblock Efficient semi-quantum private comparison using single photons.
\newblock {\em Quantum Information Processing}, 18(7):207, 2019.

\bibitem{sqpc-no-ent}
Tian-Yu Ye and Chong-Qiang Ye.
\newblock Measure-resend semi-quantum private comparison without entanglement.
\newblock {\em International Journal of Theoretical Physics},
  57(12):3819--3834, 2018.

\bibitem{sqkd-ident}
Xiao-Jun Wen, Xing-Qiang Zhao, Li-Hua Gong, and Nan-Run Zhou.
\newblock A semi-quantum authentication protocol for message and identity.
\newblock {\em Laser Physics Letters}, 16(7):075206, 2019.

\bibitem{sqkd-ident-2}
Nan-Run Zhou, Kong-Ni Zhu, Wei Bi, and Li-Hua Gong.
\newblock Semi-quantum identification.
\newblock {\em Quantum Information Processing}, 18(6):197, 2019.

\bibitem{database}
Min Xiao and Di-Fang Zhang.
\newblock Practical quantum private query with classical participants.
\newblock {\em Chinese Physics Letters}, 36(3):030301, 2019.

\bibitem{sqkd-mdi}
Jinjun He, Qin Li, Chunhui Wu, Wai~Hong Chan, and Shengyu Zhang.
\newblock Measurement-device-independent semiquantum key distribution.
\newblock {\em International Journal of Quantum Information}, 16(02):1850012,
  2018.

\bibitem{sqkd-ot}
Yu-Guang Yang, Rui Yang, He~Lei, Wei-Min Shi, and Yi-Hua Zhou.
\newblock Quantum oblivious transfer with relaxed constraints on the receiver.
\newblock {\em Quantum Information Processing}, 14(8):3031--3040, 2015.

\bibitem{sqkd-sig}
Xing-Qiang Zhao, Hua-Ying Chen, Yun-Qian Wang, and Nan-Run Zhou.
\newblock Semi-quantum bi-signature scheme based on w states.
\newblock {\em International Journal of Theoretical Physics}, pages 1--13,
  2019.

\bibitem{bounded-storage}
Ivan~B Damg{\aa}rd, Serge Fehr, Louis Salvail, and Christian Schaffner.
\newblock Cryptography in the bounded-quantum-storage model.
\newblock {\em SIAM Journal on Computing}, 37(6):1865--1890, 2008.

\bibitem{bounded-storage2}
Ivan~B Damg{\aa}rd, Serge Fehr, Louis Salvail, and Christian Schaffner.
\newblock Secure identification and qkd in the bounded-quantum-storage model.
\newblock In {\em Annual International Cryptology Conference}, pages 342--359.
  Springer, 2007.

\bibitem{noisy1}
Stephanie Wehner, Christian Schaffner, and Barbara~M Terhal.
\newblock Cryptography from noisy storage.
\newblock {\em Physical Review Letters}, 100(22):220502, 2008.

\bibitem{noisy2}
Robert Konig, Stephanie Wehner, and J{\"u}rg Wullschleger.
\newblock Unconditional security from noisy quantum storage.
\newblock {\em IEEE Transactions on Information Theory}, 58(3):1962--1984,
  2012.

\bibitem{tag-attack}
Yong-gang Tan, Hua Lu, and Qing-yu Cai.
\newblock Comment on “quantum key distribution with classical bob”.
\newblock {\em Physical review letters}, 102(9):098901, 2009.

\bibitem{tag-atk-counter}
Michel Boyer, Dan Kenigsberg, and Tal Mor.
\newblock Boyer, kenigsberg, and mor reply.
\newblock {\em Physical Review Letters}, 102(9):098902, 2009.

\bibitem{tag-atk-2}
Yu-Guang Yang, Si-Jia Sun, and Qian-Qian Zhao.
\newblock Trojan-horse attacks on quantum key distribution with classical bob.
\newblock {\em Quantum Information Processing}, 14(2):681--686, 2015.

\bibitem{CV1}
Stefano Pirandola, Stefano Mancini, Seth Lloyd, and Samuel~L Braunstein.
\newblock Continuous-variable quantum cryptography using two-way quantum
  communication.
\newblock {\em Nature Physics}, 4(9):726, 2008.

\bibitem{CV2}
Carlo Ottaviani and Stefano Pirandola.
\newblock General immunity and superadditivity of two-way gaussian quantum
  cryptography.
\newblock {\em Scientific reports}, 6:22225, 2016.

\bibitem{CV3}
Christian Weedbrook, Carlo Ottaviani, and Stefano Pirandola.
\newblock Two-way quantum cryptography at different wavelengths.
\newblock {\em Physical Review A}, 89(1):012309, 2014.

\bibitem{CV4}
Carlo Ottaviani, Stefano Mancini, and Stefano Pirandola.
\newblock Two-way gaussian quantum cryptography against coherent attacks in
  direct reconciliation.
\newblock {\em Physical Review A}, 92(6):062323, 2015.

\bibitem{CV5}
Quntao Zhuang, Zheshen Zhang, Norbert L{\"u}tkenhaus, and Jeffrey~H Shapiro.
\newblock Security-proof framework for two-way gaussian
  quantum-key-distribution protocols.
\newblock {\em Physical Review A}, 98(3):032332, 2018.

\bibitem{CV6}
Shouvik Ghorai, Eleni Diamanti, and Anthony Leverrier.
\newblock Composable security of two-way continuous-variable quantum key
  distribution without active symmetrization.
\newblock {\em Physical Review A}, 99(1):012311, 2019.

\bibitem{CV-floodlight}
Quntao Zhuang, Zheshen Zhang, Justin Dove, Franco N.~C. Wong, and Jeffrey~H.
  Shapiro.
\newblock Floodlight quantum key distribution: A practical route to
  gigabit-per-second secret-key rates.
\newblock {\em Phys. Rev. A}, 94:012322, Jul 2016.

\bibitem{sqkd-mirror}
Michel Boyer, Matty Katz, Rotem Liss, and Tal Mor.
\newblock Experimentally feasible protocol for semiquantum key distribution.
\newblock {\em Physical Review A}, 96(6):062335, 2017.

\bibitem{sqkd-reflect-prac}
Walter~O Krawec.
\newblock Practical security of semi-quantum key distribution.
\newblock In {\em Quantum Information Science, Sensing, and Computation X},
  volume 10660, page 1066009. International Society for Optics and Photonics,
  2018.

\bibitem{sqkd-exp1}
Pavel Gurevich.
\newblock {\em Experimental Quantum Key Distribution with Classical Alice}.
\newblock Technion-Israel Institute of Technology, Faculty of Computer Science,
  2012.

\bibitem{sqkd-mirror-simple}
Michel Boyer, Rotem Liss, and Tal Mor.
\newblock Attacks against a simplified experimentally feasible semiquantum key
  distribution protocol.
\newblock {\em Entropy}, 20(7):536, 2018.

\bibitem{QKD-B92-USD}
Kiyoshi Tamaki, Masato Koashi, and Nobuyuki Imoto.
\newblock Security of the bennett 1992 quantum-key distribution protocol
  against individual attack over a realistic channel.
\newblock {\em Physical Review A}, 67(3):032310, 2003.

\bibitem{passive-switch}
GP~Temporao.
\newblock Passive switching scheme for two-way quantum key distribution setups.
\newblock {\em Electronics Letters}, 46(7):512--513, 2010.

\end{thebibliography}

\end{document}